\documentclass[a4paper,fleqn,usenatbib]{mnras}

\usepackage{txfonts}

\usepackage[T1]{fontenc}
\usepackage{ae,aecompl}

\usepackage{graphicx}	
\usepackage{amstext}
\usepackage{amssymb}	

\newcommand{\HI}{$\rm{H\,{\sevensize I}}$}

\newcommand{\lya}{Ly$\alpha$}
\newcommand{\Lya}{Ly$\alpha$}

\newcommand{\kms}{$\rm km~s^{-1}$}

\newcommand{\sbline}{$\rm erg~s^{-1}~cm^{-2}~arcsec^{-2}$}

\newcommand{\flline}{$\rm erg~s^{-1}~cm^{-2}$}
\newcommand{\CIV}{\mbox{C\,{\sc iv}}}
\newcommand{\CIII}{\mbox{C\,{\sc iii}]}}

\newcommand{\OVI}{\mbox{O\,{\sc vi}}}
\newcommand{\HeII}{\mbox{He\,{\sc ii}}}

\title[MAGG. III. The environment of $z=3-4.5$ quasars]{MUSE Analysis of Gas around Galaxies (MAGG) - III: The gas and galaxy environment of $z=3-4.5$ quasars.}

\author[Fossati et al.]{M. Fossati$^{1}$ \thanks{E-mail: matteo.fossati@unimib.it}, 
M. Fumagalli$^{1}$,
E.K. Lofthouse$^{1}$,
R. Dutta$^{2,1}$, 
S. Cantalupo$^{1,3}$, \and
F. Arrigoni Battaia$^{4}$,
J.P.U. Fynbo$^{5}$,
E. Lusso$^{6}$,
M.T. Murphy$^{7}$,
J.X. Prochaska$^{8}$,
T. Theuns$^{2}$, \and
R.J. Cooke$^{2}$
\\
  $^{1}$Dipartimento di Fisica G. Occhialini, Universit\`a degli Studi di Milano-Bicocca, Piazza della Scienza 3, 20126 Milano, Italy \\
  $^{2}$Institute for Computational Cosmology and Centre for Extragalactic Astronomy, Durham University, South Road, Durham, DH1 3LE, UK \\
  $^{3}$Department of Physics, ETH Zurich, Wolfgang-Pauli-Strasse 27, CH-8093 Zurich, Switzerland \\
  $^{4}$Max-Planck-Institut f\"ur Astrophysik, Karl-Schwarzschild-Str 1, D-85748 Garching bei M\"unchen, Germany \\
  $^{5}$Niels Bohr Institute, University of Copenhagen, Lyngbyvej 2, DK-2100 Copenhagen, Denmark \\
  $^{6}$Dipartimento di Fisica e Astronomia, Universit\`a degli Studi di Firenze, Via Giovanni Sansone, I-50019 Sesto Fiorentino, Italy \\
  $^{7}$Centre for Astrophysics and Supercomputing, Swinburne University of Technology, Hawthorn, Victoria 3122, Australia \\
  $^{8}$Department of Astronomy and Astrophysics, University of California, Santa Cruz, CA 95064, USA \\
 }

\date{Accepted 2021 March 01. Received 2021 February 01; in original form 2020 December 02}

\pubyear{\the\year}

\begin{document}
\label{firstpage}
\pagerange{\pageref{firstpage}--\pageref{lastpage}}
\maketitle

\begin{abstract}

We present a study of the environment of 27 $z=3-4.5$ bright quasars from the MUSE Analysis of Gas around Galaxies (MAGG) survey. With medium-depth MUSE observations (4 hours on target per field), we characterise the effects of quasars on their surroundings by studying simultaneously the properties of extended gas nebulae and \lya\ emitters (LAEs) in the quasar host haloes.
We detect extended (up to $\approx 100$~kpc) Ly$\alpha$ emission around all MAGG quasars, 
finding a very weak redshift evolution between $z=3$ and $z=6$.
By stacking the MUSE datacubes, we confidently detect extended emission of \CIV\ and only marginally detect extended \HeII\ up to $\approx 40 $~kpc, implying that the gas is metal enriched.  
Moreover, our observations show a significant overdensity of LAEs within 300 \kms\ from the quasar systemic redshifts estimated from the nebular emission. The luminosity functions and equivalent width distributions of these LAEs show similar shapes with respect to LAEs away from quasars suggesting that the \lya\ emission of the majority of these sources is not significantly boosted by the quasar radiation or other processes related to the quasar environment. Within this framework, the observed LAE overdensities and our kinematic measurements imply that bright quasars at $z=3-4.5$ are hosted by haloes in the mass range $ \approx 10^{12.0}-10^{12.5}~\rm M_\odot$. 
\end{abstract}




\section{Introduction} \label{sec:intro}
Since the early 2000s, astronomical observations coupled to theoretical models have built an increasingly realistic and sophisticated framework for galaxy formation and evolution, laying its foundations on the cosmological
$\Lambda$ Cold Dark Matter ($\Lambda$CDM) model.  Galaxies form within dark matter haloes that grow hierarchically \citep{Gunn72, White78, Perlmutter99}, and are shaped in their morphology and physical properties by several, and often competing, processes that regulate their evolution, giving rise to the diverse galaxy population we observe in the present-day Universe. Despite the advancements both in the observational and theoretical side, the complex interplay of physical processes taking place during the early stages ($z\gtrsim 3$) of galaxy formation and halo assembly remains an open question in modern galaxy formation theories.

With the exception of the most massive galaxies, the primary mode of galaxy growth is via the star formation process \citep{van-Dokkum13, Wilman20}, that converts gas into new stellar populations. This process is tightly related to the balance of gas accretion into the interstellar medium (ISM) and outflows \citep{Bouche10, Lilly13, Sharma20}. Gas can be acquired through the cooling of a hot halo \citep{White91} or the inflow through cold gas streams \citep{Keres05, dekel06, Dekel09, van-de-Voort11, Theuns21}, while feedback processes due to supernovae or active galactic nuclei (AGN) are deemed responsible for gas ejection back into the halo around galaxies: the circum-galactic medium (CGM). The CGM gas, which is inherently multi-phase, has a tight link to the demographics and properties of early galaxies and is therefore a key piece in the puzzle of galaxy formation. 

The physical properties of the CGM, predicted to be diffuse and difficult to detect in emission, have been first studied from spectral absorption features, using bright background sources as light probes \citep[e.g.][]{Bergeron02, Hennawi07, Steidel10, Rubin10, Prochaska11, Rudie12, Fumagalli13, Fumagalli16a, Tumlinson13, Tumlinson17, Bordoloi14, Turner14, Turner17a}. Although powerful in reaching very low density gas, this technique has its main limitation in its sparse spatial coverage, as a single sightline gives only a point source estimate of the probed gas.

Several techniques have been exploited to attempt to overcome this problem, including the use of quasar pairs \citep[e.g.][]{Martin10}, lensed
quasars \citep[e.g.][]{Chen14,Rubin18b}, giant lensed arcs \citep[e.g.][]{Lopez18}, or massive galaxies \citep[e.g.][]{Rubin18} acting as multiple background sources at small projected separations. Moreover, with experiments observing quasar pairs (e.g. the Quasars Probing Quasars, QPQ \citealt{Hennawi07}, or similar ones \citealt{Bowen06, Farina13}) it became possible to probe the CGM of bright quasars in two orthogonal directions (i.e. along the line-of-sight and transverse), paving the road for tomographical studies of the CGM in absorption. 

Conversely, studies of the CGM in emission have traditionally been spatially resolved and have focused on \lya\ emitting gas near high-redshift quasars, where emission was detected at radii $R<50$ kpc in $>50\%$ of the objects \citep{Hu87, Fynbo99, Weidinger04, Weidinger05, Christensen06, Hennawi13}. More recently, deeper observations with a variety of observing techniques (long slit spectroscopy, narrow band imaging, integral field spectroscopy) revealed even more extended ($R>100$ kpc) \lya\ nebulosities around bright quasars \citep[][]{Cantalupo14, Hennawi15, Arrigoni-Battaia18a, Arrigoni-Battaia19, Cai19, Farina19}.

A revolution in the study of the CGM in emission has been the development of new, highly sensitive, integral field spectrographs (IFS) mounted on  $8-10~$m ground based telescopes, like the Multi Unit Spectroscopic Explorer \citep[MUSE,][]{Bacon10} at the ESO Very Large Telescope (VLT), and the Keck Cosmic Web Imager \citep[KCWI,][]{Morrissey18}. These instruments provide a deep spatially resolved view of the CGM in emission reaching unprecedented faint surface brightness levels (e.g. $10^{-18}$ \sbline\ for emission lines) with reasonable observing times ($5-10~$h). Observations of the CGM around $z\sim2-5$ quasars became routine with these instruments, contributing to the build-up of larger and unbiased samples \citep{Borisova16, Marino19, Cai19, Arrigoni-Battaia19, Farina19}. Among these, large and shallower surveys found that extended nebulae are ubiquitous around quasars.

At the same time, deeper IFS observations opened up new possibilities in  the study of the connection of quasar hosts and their halo environment. In particular, low mass galaxies at $z>3$ are expected to be experiencing rapid growth in a gas rich and almost optically dark phase \citep{Dekel09}, as deep ALMA observations are starting to reveal \citep{Franco18}. \citet{Cantalupo12}, using narrow-band imaging techniques, reported the discovery of nearly 100 candidate dark galaxies at $z=2.4$ brought to light by the \lya\ fluorescence induced by a nearby hyperluminous quasar. More recently, \citet{Marino18} and \citet{Li19} showed the potential of very deep IFU observations with MUSE or KCWI to study these dark galaxies finding candidate populations near several quasars at $z\sim 3-3.5$. 

Still, a clear picture regarding the local galaxy environment of $z>3$ quasars remains partially contradictory. The quasar auto-correlation function \citep{Myers07, Shen07, Eftekharzadeh15, He18, Timlin18} shows that, at least in the redshift range $0.5<z \lesssim 4$, quasars are a highly clustered population, possibly living in massive dark matter haloes for their epoch ($M_h \approx 10^{12.5}{\rm M}_\odot$). Within this framework, \citet{Hennawi06} found a significant evidence that the quasar auto-correlation function gets even steeper on scales below a megaparsec. However, it remains unclear if quasars reside in overdensities of galaxies. Many works have indeed attempted to characterize the density of galaxies around quasars (mostly through clustering analysis), finding either a galaxy overdensity \citep{Kashikawa07, Utsumi10, Garcia-Vergara17, Garcia-Vergara19, Mignoli20} or a number density consistent with field samples \citep{Toshikawa16, Mazzucchelli17, Uchiyama19}. At $z\sim5-6$ there is growing evidence that quasars can be found in overdensities of emission line or submillimeter galaxies \citep{Farina17, Decarli17, Trakhtenbrot17}. A definitive interpretation of the different results found so far is complicated by the various methods used to identify galaxies (leading to inhomogeneous populations), by the fluctuations arising from cosmic variance in small quasar samples, and by the different spatial scales probed by these works.

The MUSE Analysis of Gas around Galaxies (MAGG) survey \citep{Lofthouse20} is a large and medium-deep MUSE survey covering 28 fields centered on $z=3.2-4.5$ quasars. The MUSE data are primarily from our VLT large programme (ID 197.A$-$0384, PI: M. Fumagalli), and are supplemented by data from the MUSE GTO \citep[PI: J. Schaye;][]{Muzahid20}. The program is complemented by sensitive high resolution spectroscopic observations of the quasars taken with instruments mounted at VLT, Keck, and Magellan telescopes. The survey strategy and methodology, including sample selection and data processing, have been presented in \citet{Lofthouse20}. 
The main goal of MAGG is the study of the CGM of $z=3-4.5$ star-forming galaxies in the surroundings of  \HI\ absorbers with $\log N_{\rm HI} \gtrsim10^{17}~\mathrm{cm^{-2}}$ (Lofthouse et al. in prep.). The rich MAGG datasets are further being utilised to study the cold CGM gas around $z\sim1$ galaxies \citep{Dutta20}, and near $z=3-4.5$ quasars, which is the subject of this work.

One key element of novelty of the MAGG survey, compared to other published studies of the quasar environment is its combination of medium depth (4~h on source per field) and the large number of independent fields (28), which places MAGG in between larger but shallower surveys \citep{Borisova16, Arrigoni-Battaia19, Farina19} and deeper surveys ($\gtrsim10~$h) of a smaller number of fields \citep{Marino18, Bacon17, Lusso19, Fossati19a}. 
Thanks to these unique characteristics, the MAGG survey is ideal to coherently study the CGM and the halo environment of high-redshift quasars with a large, yet deep sample.
In this paper we characterize the properties of the nebulae around the quasar hosts with a focus on the metal content and density of the CGM. We also study the population of \lya\ emitters (LAEs) in the vicinity of the quasars, focussing on their spatial and luminosity distributions and on the spatial and spectral properties of their CGM. Throughout this work, we will also compare the properties of LAEs with those found near high-column density absorbers and in the field, to find key similarities and differences.

The paper is structured as follows: in Section~\ref{sec:data} we describe the MAGG dataset and we summarize the data reduction steps. In Sections~\ref{sec:detnebulae} and \ref{sec:detLAE} we present the algorithms we use to identify both extended ionized gas nebulae and compact LAEs. Our results are presented in Section~\ref{sec:results}, and we discuss them in the context of the co-evolution of gas and galaxies in the environment of high-redshift quasars in Section~\ref{sec:discussion}. Our conclusions are summarized in Section~\ref{sec:conclusions}.

Throughout this paper, we assume a flat $\Lambda$CDM cosmology with $H_0 = 67.7~{\rm km~s^{-1}~Mpc^{-1}}$ and $\Omega_m = 0.307$ \citep{Planck16}, and all magnitudes are expressed in the AB system.

\section{Data} \label{sec:data}
The MAGG survey is built upon a MUSE Large Programme (ID 197.A-0384; PI Fumagalli) of 28 quasar fields at $z \approx 3.2-4.5$ for which high-resolution spectroscopy of the quasars is available. A complete description of the target selection, the quasar spectroscopy and the data acquisition and reduction techniques is given in \citet{Lofthouse20} and these details are  only briefly summarized here. 

The MAGG sample comprises 28 quasars with $m_r<19$ mag, with archival high-resolution spectroscopy at $S/N>20$, each one with at least one strong hydrogen absorption line system ($N_{\rm HI} > 10^{17}~{\rm cm^{-2}}$) at redshift $z > 3.05$. We also require a position in the sky that is observable from VLT with low airmass, corresponding to a declination range $-40$~deg $ < \delta < 15$~deg. Each quasar field has been observed with MUSE between ESO periods 97 and 103 for a total on-source time of $\approx4$ h per field, with longer exposure times in fields with partial MUSE observations from the archive. The MUSE observations include dithers and instrument rotations of 90 deg to improve on the flatness of the field, thus mitigating the differences in the performance of the MUSE spectrographs. All the exposures have been taken on clear nights, at airmass $<$ 1.6, and with an average image quality of 0.6-0.7 arcsec full width at half maximum (FWHM). 

The MUSE raw data are first reduced with the ESO MUSE pipeline \citep[v2.4.1,][]{Weilbacher14}, to remove instrumental signatures from the data applying a bias and flat-field correction and the wavelength and flux calibrations. Cubes are then reconstructed after sky subtraction and registered to the Gaia DR2 astrometry \citep{Gaia18}. Upon stacking, these data show imperfections
arising from residuals of the illumination of the detectors, and imperfections in the subtraction of sky lines. Several tools exist to mitigate these imperfections, and in MAGG we use primarily the {\sc CubExtractor} package \citep[][and Cantalupo in prep.]{Cantalupo19}.

The {\sc CubEx} processing starts from resampling each non sky subtracted pixel table into a datacube on a fixed reference grid that is derived from stacking the ESO products. Next, we use the {\sc CubeFix} tool to flatten the illumination of the field and of individual slices, and the {\sc CubeSharp} tool for a local and more accurate sky subtraction which takes into account spatial variations in the instrument line spread function. These tools are applied three times, each time refining the mask of the sources, therefore achieving a better and better correction. However, these algorithms cannot easily correct variations across individual slices. These variations are particularly prominent at the edges of the slices. For this reason, to obtain clean coadds we mask the first and the last two pixels of each slice in individual exposures. Lastly, we co-add the exposures with mean and median statistics and we also generate two coadds containing only one half of all the exposures each, which are used to identify uncorrected artefacts, such as residual cosmic rays. Unless otherwise specified, the results presented in this paper are based on the mean coadds.

The uncertainty associated with individual pixels is propagated across the various steps of the reduction including the non-linear interpolation procedure used to resample the data. However, the uncertainty in the final coadds does not accurately reproduce the effective standard deviation of the voxel (volumetric pixels) inside the final data cube. We therefore proceed to bootstrap pixels in individual exposures to estimate the noise in each of our final data products \citep[see also][for a complete description of the procedure]{Fossati19a}. Due to the small number of individual exposures, we then scale the pipeline variance cube with a wavelength-dependent function obtained from the bootstraps. The final result is a series of datacubes with accurate standard deviations, which are required for a robust detection of sources.

\begin{figure*}
    \centering
    \includegraphics[width=0.90\textwidth]{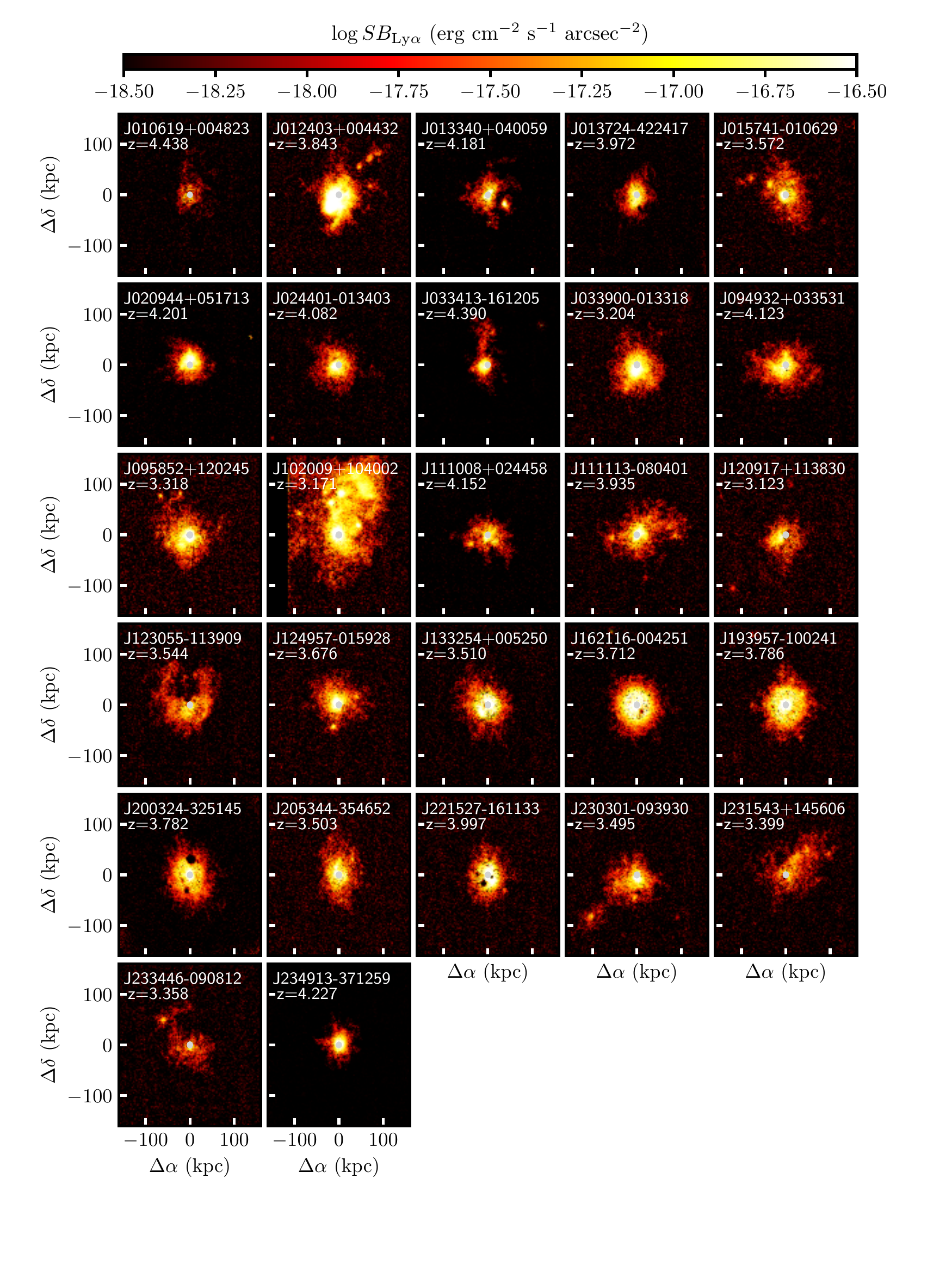}
    \caption{Observed surface brightness maps of extended \lya\ nebulae detected around the 27 MAGG quasars included in our study, and sorted by right ascension. The spatial scale is in proper kpc. The grey circle at the center of each image marks a 10 kpc radius where the quasar PSF residuals dominate the signal. }
    \label{fig:nebgallery}
\end{figure*}

\section{Detection of extended nebulae around quasars} \label{sec:detnebulae}

The detection of extended nebulae surrounding quasars requires further processing of the data, most notably the removal of the quasar point spread function (PSF) and other compact continuum or line emitting sources in the field. For these tasks we use tools from the {\sc CubEx} package, similarly to what has been done by \citet{Borisova16} and \citet{Arrigoni-Battaia19}, as detailed below.

We first subtract the PSF of the quasar from each combined cube  using the {\sc CubePSFSub} tool. This code first generates a set of PSF images from narrow band images obtained in bins of 250 MUSE spectral pixels. This ensures that the wavelength dependence of the seeing FWHM is included in the model. Then, the PSF image is rescaled to match the quasar flux within a $1\times1~$arcsec$^2$ region for each wavelength, which is then subtracted from the data. As stated in \citet{Arrigoni-Battaia19}, this choice makes the assumption that the quasar is much brighter than the host galaxy within the rescaling area and makes the inner 1~arcsec of the data unusable for science.  However, we tested that this method has superior performance compared to a PSF model estimated from stars in the field because small variations of the PSF across the MUSE field leave strong residuals in the  quasar PSF subtraction. We run the same tool on stars in the field, which we identify from their spectral shape using the MARZ tool as described in \citet{Lofthouse20}. 

Then, we subtract all the remaining continuum sources which cannot be modelled as point sources using the {\sc CubeBKGSub} tool whose algorithm is described in \citet{Borisova16} and \citet{Cantalupo19}. We are then left with a set of cubes free from any continuum source, which can be used to search both for \lya\ emitters (as described in Section \ref{sec:detLAE}) and for extended nebulae around the quasars. 

To extract the extended \lya\ emission we run {\sc CubeExtractor} on a portion of the cube (300 spectral pixels wide) around the quasar systemic redshift. We filter the cube with a boxcar spatial filter of 2 pixels before detection to increase the coherence of the detected signal. We initially define the nebulae to be sources with at least 10000 connected voxels with individual signal to noise $S/N>2$ and with a geometrical center within 10~arcsec of the quasar position. The distance constraint rejects spurious sources at the edges of the cube where the noise estimates are more uncertain. With this setup, we detect extended \lya\ emission around most quasars. However, some emission is too faint to satisfy the 10000 voxels volume, we therefore reduced this threshold in steps of 1000 until a single detection is eventually found. All our nebulae cover a volume of at least 5000 voxels. We further tested the reliability of the detection method by running the same algorithm on the median cubes, finding in all cases sources with a consistent 3D position in the cube, similar morphology, and comparable total flux. 

We note that the quasar J142438+225600 included in the original MAGG sample is lensed by a $z\approx 0.34$ source \citep{Patnaik92}, leading to a complex morphology of the \lya\ emission in the image plane. For this reason, we exclude it from this analysis, leaving a dedicated modelling of the lens to a future publication. This paper therefore focuses on a total of 27 fields, out of the 28 in the MAGG sample.
Figure \ref{fig:nebgallery} shows a gallery of the optimally extracted \lya\ emission around the 27 MAGG quasars we study here, which we obtain by collapsing the voxels within the {\sc CubEx} detection region (usually defined as the segmentation cube) along the spectral direction with empty pixels filled with data from the MUSE wavelength layer corresponding to the peak of the \lya\ emission (consistently with the method adopted by \citealt{Borisova16} and \citealt{Arrigoni-Battaia19}).

After detecting extended \lya\ emission we extract a spectrum from the spatial pixels that appear in the 3D segmentation cube. We define the redshift of the nebula to correspond to the wavelength of the peak of the \lya\ emission in the spectrum. The systemic redshifts of our quasars are mostly obtained from blue-shifted lines leading to possible offsets compared to the \lya\ emission of the nebulae. In what follows we will use the redshift of the nebulae $z_{\rm neb}$ when computing velocity offsets and other redshift dependent quantities.

\section{Detection of compact \lya\ emitters} \label{sec:detLAE}

The MUSE integral field data allow us to study not only the most immediate regions surrounding the quasars, but also the possible presence of star forming galaxies at larger distances up to $\sim 200~\rm kpc$ (LAEs). This is particularly possible thanks to the depth of MAGG data, which are $\approx 4\times$ deeper than data from previous studies (as discussed in Section \ref{sec:intro}) that focused exclusively on the properties of the quasar nebulae. 

\subsection{Detection and visual inspection}
We extract candidate line emitters following the procedure first described in \citet{Fumagalli16}, and later updated in \citet{Lofthouse20}. First we run {\sc CubEx} on the continuum and PSF subtracted cubes after masking continuum-detected sources of known redshift (including stars). The extraction is similar to the one used for the extended nebulae but with different thresholds, such that candidate emitters need to have: (i) a segmentation cube  of more than 27 voxels, (ii) pixels covering at least 3 wavelength channels ($>3.7\AA$) along at least a spatial pixel, (iii) a global segmentation cube not spanning more than 20 wavelength channels, to optimally reject continuum source residuals. These selection constraints are quantitatively driven by the instrument spatial point spread function and the spectral line spread function in order to separate candidate emitters from artefacts in the data (which are unlikely to be more extended than a few pixels either spatially or spectrally). We verified that no additional real source is found if constraint (iii) is extended to 40 wavelength channels.

The candidate line emitters are then classified into two confidence levels based on their integrated $S/N$ ($ISN$), corrected for the noise covariance as described in detail in \citet{Lofthouse20}. Class 1 sources are characterized by $ISN>7$ in the mean combine, while class 2 includes sources with $5<ISN<7$ which extends the completeness of the sample at the expenses of larger uncertainties on the photometry and redshift, and to some extent of the sample purity. For a given field, candidates within $\pm 1000$ \kms\ from the redshift of the nebulae are visually inspected by three authors to confirm their LAE nature. To probe different environments (see Section \ref{sec:LF}) we also compare our LAE sample to LAEs identified within $\pm 1000$ \kms\ of the redshift of strong hydrogen absorbers, which have been selected and processed using the same procedures described here. Cosmic rays are the  most common source  of contamination in the LAEs catalogue. To identify them we monitor the $ISN$ in the median and half exposure coadds. In these metrics, cosmic rays appear as high $S/N$ sources only in a single half exposure coadd, which in turn affect the $ISN$ of the median coadd.  During the visual inspection we also look at the shape of the segmentation cube and the extracted emitter spectrum to identify possible skyline residuals or other artefacts that might be present near the edges of the MUSE field of view.

\begin{figure*}
    \centering
    \includegraphics[width=0.95\textwidth]{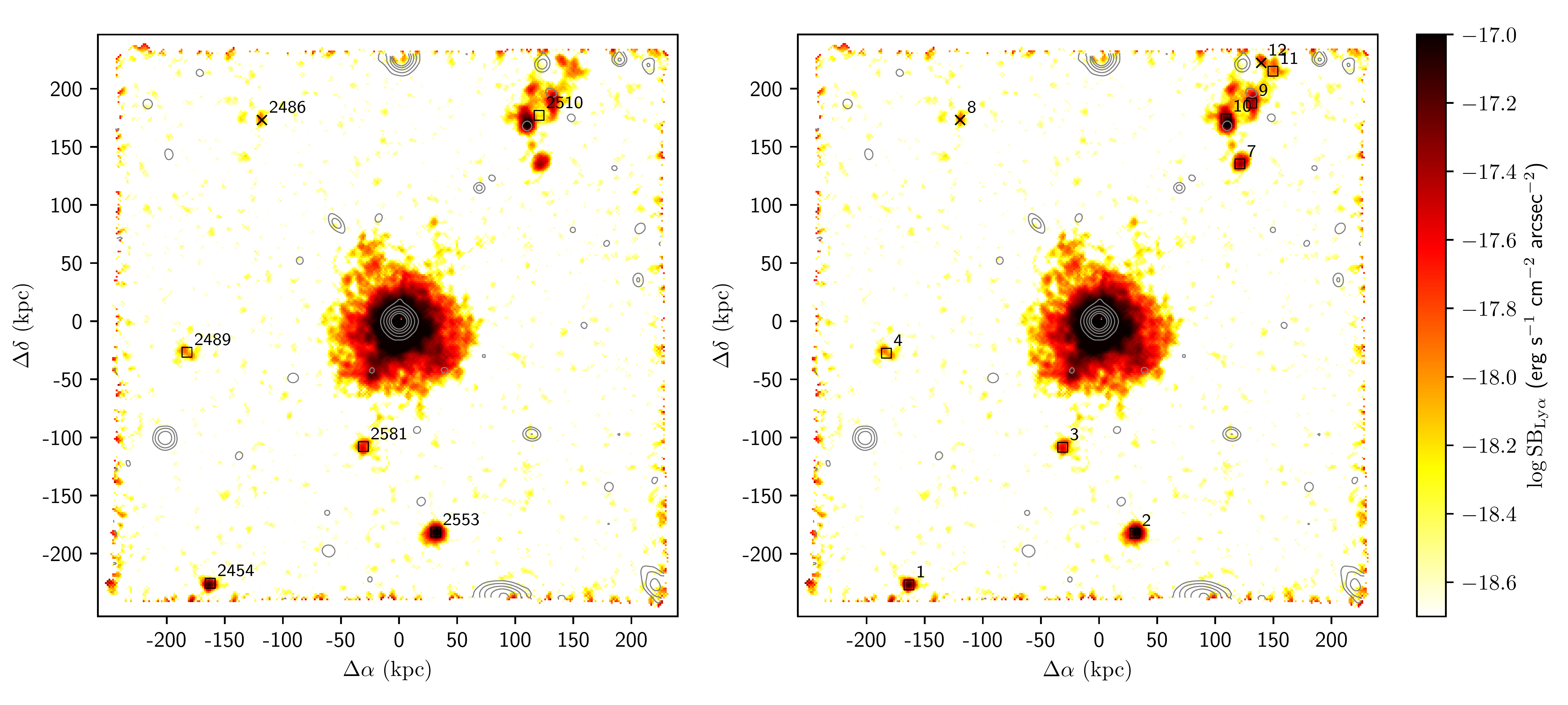}
    \caption{Left: A reconstructed image of the \lya\ emission at the redshift of the extended nebula for the field J033900$-$013318. The image is obtained by combining the optimally extracted flux maps for the compact emitters and the extended nebula and with the sum of four wavelength channels centered at $z_{\rm neb}$ elsewhere. The image is smoothed with a top-hat kernel of width $0.4~$arcsec (equal to 2 MUSE pixels). The gray contours show the continuum sources, and are uniformly spaced between 22 and 27 mag arcsec$^{-2}$. The black squares mark the position of detected class 1 LAEs, while the black crosses mark lower confidence class 2 sources. Their catalogue ID is also shown.
    Right: same as left panel but with LAEs identified after our deblending procedure. Multiple clustered emitters are separated into individual components. }
    \label{fig:J033_emitters}
\end{figure*}

\subsection{Source deblending}

The left panel of Figure \ref{fig:J033_emitters} shows, as an example, the extended \lya\ nebula and the LAEs identified in the field J033900$-$013318. While the vast majority of the LAEs are isolated compact sources, in some cases they appear to be clustered such that different emitters are connected into a single source by the {\sc CubEx} algorithm (see e.g. ID 2510 in Figure \ref{fig:J033_emitters}). The {\sc CubEx} code can be run with a deblending option, however, because our master catalog has been run only once for the MAGG programme and is shared for different science goals, in this work we employ a specific procedure to separate clustered emitters. First we generate a composite image of the \Lya\ emission as shown in Figure \ref{fig:J033_emitters}, then we run {\sc SExtractor} \citep{Bertin96} masking all the pixels not belonging to the nebula and the LAEs, and we deblend the sources using a {\tt DEBLEND\_CONT} parameter of 0.05. This value has been chosen after extensive testing because it optimally separates clustered LAEs and selects bright and compact sources in the outer regions of the nebulae without over-shredding the extended \lya\ signal into spurious sources. Every source identified in the {\sc SExtractor} run must also have an $ISN>5$ and a minimum size of 9 pixels to be retained in the final catalogue. The right panel of Figure \ref{fig:J033_emitters} shows how the source ID2510 has been split into five sources, four of them being high-confidence class 1 LAEs and one being a lower confidence source possibly due to the higher noise near the edge of the field of view (FoV). For each deblended source we generate an appropriate 3D segmentation cube and we extract a spectrum from the spatial pixels where it is identified. 

Combining the above procedures, we find a total of 113 LAEs in our 27 fields, out of which 85 are class 1 sources. 
As a last step, we compute the emitter redshift from the peak of its \lya\ emission in the spectrum. For the emitters that exhibit a double peaked \lya\ or a blue bump, we estimate the redshift from the wavelength corresponding to the average of the two peaks. In these cases, \citet{Verhamme18} have shown this average value is in excellent agreement with the redshift estimates obtained from non-resonant lines.

\subsection{Total flux estimate of the \lya\ line}
It has been recently shown \citep{Wisotzki16,Leclercq17} that the spatial emission of LAEs can typically be decomposed in a bright core and a faint diffuse halo. As a result, the emitter flux given by {\sc CubEx} in the segmentation cube typically underestimates the total flux by missing the faint and diffuse emission. 
Following \citet{Marino18}, we compute total \lya\ fluxes from a curve of growth (CoG) analysis. For each emitter we generate a pseudo narrow-band (NB) image by summing the spectral channels within $\pm 15~\AA$ from its redshift. After masking neighbouring sources, we generate a flux CoG in circular apertures spatially centered on the {\sc CubEx} coordinates and with radii increasing in steps of $0.2~$arcsec. We test on our brightest emitters that the largest radius does not exceed $3~$arcsec and therefore we set this maximum value for all emitters. 

We perform a local background subtraction by taking the median surface brightness in a concentric circular annulus with inner and outer radii of $3~$arcsec and $4~$arcsec respectively and by subtracting this value, scaled to the size of each aperture, from the CoG fluxes. 
The total \lya\ flux of each object is then assumed to be the CoG flux at the last radius where the total flux grows by more than 2.5\% with respect to the previous step. We  visually inspect all CoG diagnostic plots to ensure that the flux is dominated by the \lya\ emission. Where necessary, we further mask artefacts and field edges. Lastly, we convert fluxes into luminosities using the luminosity distance at the redshift of the LAE and our chosen cosmological model.

\subsection{The LAE selection function} \label{sec:selectionfunc}
To study the statistical properties of the LAEs in the quasar environment, we need to characterize the selection function of LAEs in the MAGG survey. To assess the probability of finding a real LAE at a given flux (or luminosity) and redshift in our data, we run simulations by injecting mock sources in our datacubes and testing the ability of the detection algorithm to retrieve them. 

We use two types of sources: first we inject model sources that are unresolved (point sources) both spatially and spectrally. We use a 3D Gaussian profile with spatial FWHM=$0.6~$arcsec and a spectral FWHM equal to the spectral Nyquist sampling. Each source has a random flux that ranges between $1\times10^{-19}$ and $8\times 10^{-17}$ \flline\ and the source is randomly placed in an available portion of the datacube. To this end, continuum sources and line emitters are masked and no artificial source can overlap with them. Unexposed pixels and the edges of the FoV are also masked. After each injection a region equal to 5 times the FWHM around its center is made unavailable for other sources to avoid superpositions in the spectral direction. For each field we inject 500 random sources and we run {\sc CubEx} to identify them, we then iterate the full procedure on the 27 fields 1000 times until we obtain a statistically robust sampling of the flux, redshift and spatial positions of mock sources.

\begin{figure*}
    \centering
    \includegraphics[width=0.95\textwidth]{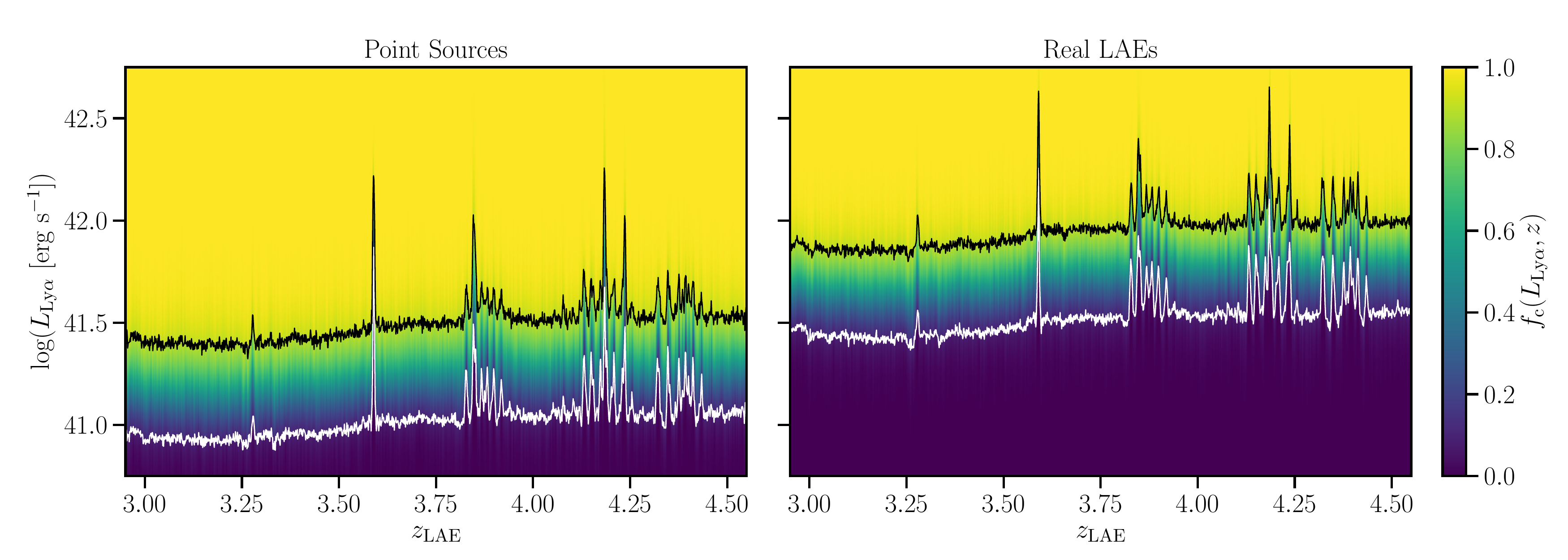}
    \caption{Fraction of detected mock LAEs (selection function) in the datacubes of the MAGG survey as a function of redshift and emitter luminosity for point source models (left panel) and real LAE models (right panel). The white and black lines mark the 10\% and 90\% detection fractions, respectively. }
    \label{fig:selfun}
\end{figure*}

We then repeat this experiment injecting real LAEs to better represent the morphology of the sources. Following \citet{Herenz19} who proposed to take these sources from deep MUSE observations to preserve the instrument sampling without the need to further process the data, we use sources from our MUSE Ultra Deep Survey \citep[MUDF,][]{Fossati19a}. Observations in the MUDF have reached, at the time of writing, a total exposure time of 70~h. By selecting 13 LAEs with $ISN>25$, we obtain data that are almost noise-free compared to the MAGG data, even when scaled to the highest flux level in our range. For each selected MUDF source, we extract a cube cutout of $29\times29$ spatial pixels and 19 spectral pixels and we normalize the data to the total CoG flux. The MUDF LAEs are selected at $3.0<z<4.4$ to cover a range of morphologies, from compact to more extended. As such, they are representative of the population of LAEs we detect in the MAGG survey in the same redshift range. The MUDF image quality is similar to the one in MAGG and therefore we do not apply any smoothing in the spatial direction. We inject these real sources by randomly selecting them and placing them in the available regions of MAGG cubes scaling them to the same flux levels we used for the point sources. 

After running these simulations, we analyse them by evaluating the fraction of recovered sources in all fields in bins of flux and redshift, $f_c(f,z)$. We then convert fluxes into luminosities in each redshift bin to obtain our final selection function $f_c(L_{\rm Ly\alpha},z)$, which is shown in Figure \ref{fig:selfun} both for the mock point sources (left) and the real sources (right). In these panels we restrict to the redshift range covered by our quasar environments. The redshift dependence of $f_c$ follows the datacube background noise which encodes the MUSE sensitivity function and the night sky flux (most notably the presence of atmospheric sky lines). While the selection function for point sources is an ideal limiting depth of our data, real emitters (which do not behave like point sources) provide a more appropriate description of the selection function of LAEs in our survey, and we will use this metric hereafter unless otherwise stated. We find the 10\% (90\%) completeness to be at $L\approx 10^{41.0}~ (10^{41.5})~\rm{erg~s^{-1}}$, with a weak dependence on redshift across the range studied in this work.

\section{Results} \label{sec:results}
In this section we focus on empirical results on the environment of $z=3-4$ quasars, leaving most of the physical interpretation to Section \ref{sec:discussion}. We start by describing the properties of the extended \lya\ emission, moving then to the properties of LAEs in the quasar haloes.

\subsection{Properties of the extended nebulae} \label{sec:results_neb}
Leveraging the unique combination of depth and sample size of our MAGG program, equivalent to $\approx 110$h of MUSE time, we can study the radial profiles of the \Lya\ emission as well as search for extended emission in fainter metal transitions inside the extended quasar nebulae. 

\subsubsection{The radial profile of the extended \lya\ emission}
\begin{figure*}
    \centering
    \includegraphics[width=1.0\textwidth]{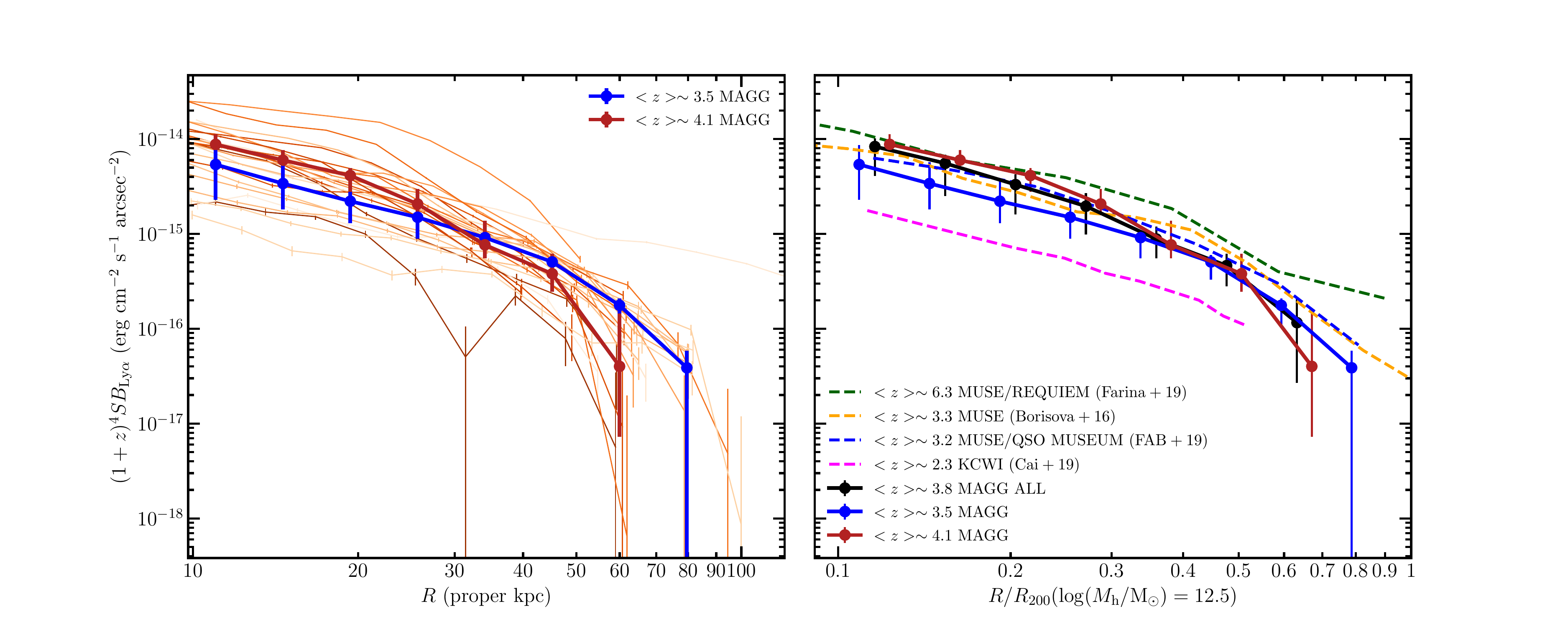}
    \caption{Left panel: the thin lines show the azimuthally averaged \lya\ profiles corrected for surface brightness dimming for the 27 extended nebulae studied in the MAGG survey. Lines are color coded by their redshift, with increasing redshift being represented by decreasing line transparency. The radial distance from the centre is expressed in proper coordinates. The blue (red) thick line shows the median profile with the error bars representing the 25$^{\rm th}$ and 75$^{\rm th}$ percentiles from the sub-sample of nebulae below (above) the median redshift, respectively. Right panel: the median \lya\ profile of all the MAGG nebulae (black solid line) and the two sub-samples at different redshifts (blue and red solid lines) normalized to the virial radius ($R_{200}$) of a $10^{12.5} {\rm M_\odot}$ dark mater halo at the average redshift of each sample. Dashed lines show the median profiles from a compilation of literature samples (see text for details) covering the redshift range $z\approx 2.2-6.3$.}
    \label{fig:neblyaprofiles}
\end{figure*}

To generate the radial profile of the \Lya\ emission we cannot use optimally extracted images, due to their truncation at $SNR>2$ per voxel which would lead to a loss of flux at low surface brightness. 
 Instead, following \citet{Arrigoni-Battaia19}, we construct narrow band images where the extended emission can be traced to lower surface brightness levels. NB images are built by summing the MUSE wavelength channels within $\pm 15~\AA$ from the peak of the nebular \lya\ emission line, and a propagated variance image is also generated. We then extract radial surface brightness profiles in circular annuli centered at the quasar coordinates. The profiles are then corrected for surface brightness dimming and are shown, for individual sources, in Figure \ref{fig:neblyaprofiles}.

 As described above, the inner region of the profile is dominated by the PSF subtraction residuals and therefore our profiles are shown only at $R>10~$kpc. 
 
 Owing to the higher redshift of MAGG quasars ($z_{\rm med} = 3.75$) compared to previous studies at $z\approx 3.2-3.3$, we can test explicitly for the presence (or lack thereof) of redshift evolution in the surface brightness profiles. To this end, we use the line color to encode redshift, with fainter line colors representing decreasing redshift. Examining how the profiles change with redshift, we observe a weak trend with higher redshift nebulae being intrinsically brighter. We further test this trend by stacking the profiles with median statistics for two sub-samples above and below the median redshift of the full MAGG sample at $z_{\rm med} = 3.75$. The median profile for higher redshift nebulae is indeed brighter then the lower redshift sub-sample, although the difference is modest. The average and median profiles for the full MAGG sample, as well as the high-z and low-z subsamples are tabulated in Table \ref{tab:MAGGprofiles}.

\begin{table*}
    \centering
    \begin{tabular}{lccccc}
    \hline
        $R$ (kpc) & Average  & r.m.s. & Median   & 25$^{\rm th}$ & 75$^{\rm th}$ \\
                  & \Lya\ SB &        & \Lya\ SB & percentile    & percentile    \\

    \hline
         11     & 81.8 (59.4,106.0)& 54.6 (42.5,55.8) & 83.4 (53.9,87.9) & 40.8 (22.9,83.4) & 101.5 (86.4,113.0) \\
         14.6   & 55.9 (38.3,74.8) & 41.8 (28.7,45.2) & 55.6 (34.0,60.0) & 25.1 (18.1,55.6) & 68.2 (57.0,76.3) \\
         19.4   & 38.1 (25.2,52.0) & 33.5 (18.7,39.8) & 33.1 (22.1,41.3) & 16.1 (12.9,27.9) & 42.5 (37.4,49.7) \\
         25.7   & 22.0 (15.6,28.9) & 20.8 (10.9,26.0) & 19.6 (15.0,20.7) & 9.9 (8.9,15.8)   & 26.8 (21.5,29.7) \\
         34.1   & 10.4 (8.6,12.2)  & 9.0 (5.3,11.5)   & 8.8  (9.1,7.7)   & 5.5 (5.5,5.5)    & 11.9 (10.4,13.8) \\
         45.2   & 4.7  (4.7,4.7)   & 3.4 (3.1,3.7)    & 4.6  (5.1,3.8)   & 2.8 (3.3,2.5)    & 6.2 (6.0,6.2) \\
         60.0   & 1.3  (1.9,0.7)   & 1.8 (2.1,1.2)    & 1.2  (1.8,0.4)   & 0.3 (1.1,0.1)    & 2.0 (2.1,1.4) \\
         79.5   & -0.1 (0.7,-0.7)  & 1.5 (1.7,0.8)    & -0.2 (0.4,-0.5)  & -0.5 (-0.1,-1.0) & 0.4 (0.6,-0.2) \\
    \hline
    \end{tabular}
    \caption{Average and median \lya\ surface brightness profiles of the extended nebulae around quasars in the full MAGG sample. The values in parenthesis refer to the low and the high redshift subsamples, respectively. The r.m.s. as well as the 25$^{\rm th}$ and 75$^{\rm th}$ percentiles of the profiles are also tabulated. Surface brightness values are in units of $10^{-16}$ \sbline\ and are corrected for cosmological surface brightness dimming.}
    \label{tab:MAGGprofiles}
\end{table*}

The question of the redshift evolution of the extended \lya\ profiles has already been debated in the literature, starting with the work by \citet{Arrigoni-Battaia19} who found that the average profile of $z\sim 3.2$  nebulae (see also \citealt{Borisova16}) was brighter than the one obtained with narrow-band observations of $z\sim2$ radio quiet quasars \citep{Arrigoni-Battaia16}. \citet{Arrigoni-Battaia19} also found a small difference in the average brightness of the profiles when their sample is split in two redshift bins ($z\sim 3.1$ and $z\sim 3.3$, with higher redshift objects being on average brighter). Since then, additional samples have been studied to extend the redshift range. At the lowest redshift, \citet{Cai19} used the Keck Cosmic Web Imager (KCWI) to observe a sample of $z\sim2.3$ quasars. At higher redshift, observations have been carried out with MUSE in a few individual quasar fields at $z\approx 5$ \citep{Ginolfi18, Bielby20, Drake20} and in a sample of 31 quasars at $z\sim6.3$ \citep{Farina19}. 
The median profiles from these surveys are shown with dashed lines in the right panel of Figure \ref{fig:neblyaprofiles} and are normalized to the virial radius ($R_{200}$) of a $10^{12.5} {\rm M_\odot}$ dark matter halo at the average redshift of each sample as done also in \citet{Farina19}. We compare these profiles with the stacked profiles of the MAGG sample which, by extending up to and beyond $z\approx 4$, bridges some of the gap between the lower-redshift ($z\lesssim 3.5$) and higher-redshift ($z\gtrsim 5$) work. All the profiles taken from the literature have been scaled to the cosmology adopted in this paper.

\begin{figure}
    \centering
    \includegraphics[width=1.05\columnwidth]{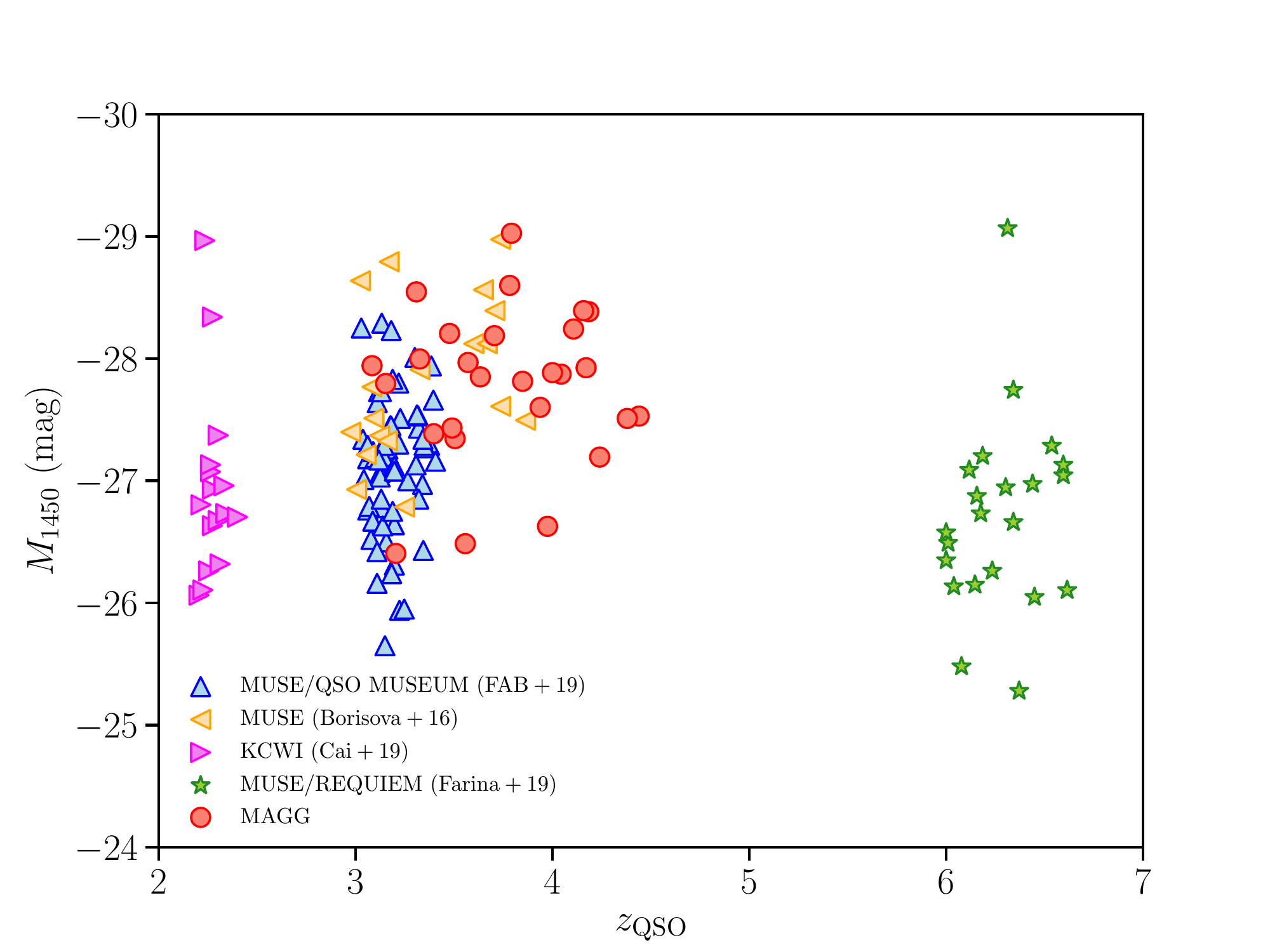}
    \caption{Redshift versus $M_{1450}$ distribution of the quasar samples with detected extended nebulae. Data from \citet{Arrigoni-Battaia19} are shown as blue triangles; from \citet{Borisova16} as orange triangles; from \citet{Cai19} as magenta triangles; from \citet{Farina19} as green stars and from our MAGG survey as red circles.}
    \label{fig:redshiftabslum}
\end{figure}

Before an interpretation of the evolution of radial profiles with redshift can be made, we need to assess whether the different quasar samples have comparable absolute luminosity. \citet{Arrigoni-Battaia19} showed that the luminosity of the \lya\ extended emission possibly scales with the luminosity of the quasar. More recently \citet{Mackenzie21}, by studying a large dynamic range in quasar luminosity, showed that the nebulae do get fainter near fainter quasars although the correlation is significantly sub-linear. As can be seen in Figure~\ref{fig:redshiftabslum}, the different literature samples we consider are composed of quasars with absolute rest-frame 1450~\AA\ magnitudes of -25 mag $<M_{1450}<$ -29 mag. The average $M_{1450}$ is consistent with -27 mag for the \citet{Arrigoni-Battaia19}, \citet{Cai19}, and \citet{Farina19}, while the \citet{Borisova16} sample has brighter quasars with $<M_{1450}>=$ -27.8 mag. Our MAGG sample is also brighter on average with $<M_{1450}>=$ -27.7 mag, but no difference is observed in the two redshift bins. Thus, no extreme variation is seen across different samples. 

\begin{figure*}
    \centering
    \includegraphics[width=0.95\textwidth]{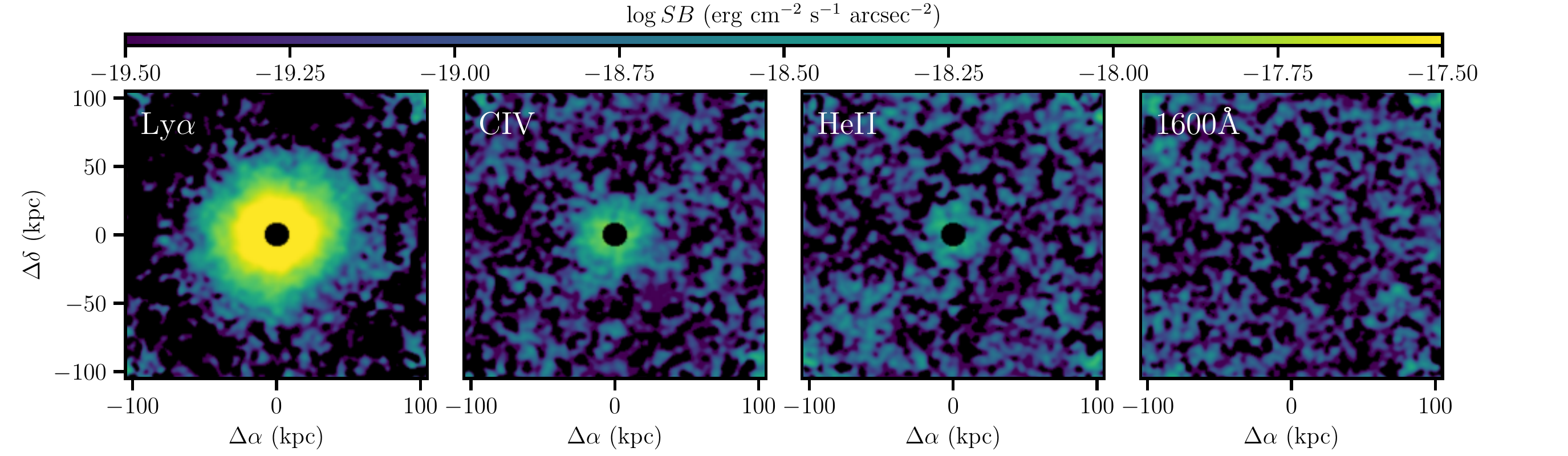}
    \caption{Two dimensional surface brightness maps of the median stack of the extended emission around the MAGG quasars. The \Lya, \CIV1549\AA, and \HeII1640\AA\ transitions are shown from left to right, as well as a stack at 1600\AA\ in the rest frame to monitor continuum subtraction residuals. The spatial axes show the proper distance from the quasar positions at the average redshift of the sample, and the black circles at the centre mask the inner 10 kpc which are dominated by the quasar PSFs.}
    \label{fig:nebstack2D}
\end{figure*}

While there appears to be a clear evolution from $z\approx 2.3$ to $z\approx 3.2-3.5$, there is only a weak redshift evolution for the profiles between $z\approx3-6$. The full MAGG sample well matches the average profiles of the QSO MUSEUM and \citet{Borisova16} samples\footnote{The median profile from the \citet{Borisova16} sample is taken from Table 2 of \citet{Marino19} where the median values are computed on a linear scale as opposed to the logarithmic scale used in \citet{Borisova16}.}, despite the average redshift being marginally higher. The high redshift MAGG subsample almost matches the REQUIEM sample for $R/R_{200}<0.3$ which could be seen as an indication of a redshift trend. However, the sub-sample size is relatively small (14 objects) and if the quasar intrinsic luminosity plays some role, we cannot rule out that the observed (small) variations in the average profiles are at least in part driven by the quasar sample and not by a clear underlying redshift evolution. The $z\sim 2.2$ sample, however, remains a significant outlier, exhibiting fainter \lya\ profiles and suggesting a strong redshift evolution of the extended \lya\ emission between $z=2$ and $z=3$. 

Our MAGG data, combined with the literature samples, corroborates the scenario where at $z>3$ cold gas is accreted onto the CGM keeping the \lya\ emission roughly constant despite the halo growth \citep{Arrigoni-Battaia19, Farina19}. At lower redshift, instead, the growth of haloes shock heats the gas reducing the mass of cold gas, leading to fainter extended \lya\ emission. 

\subsubsection{Metal line emission in the extended nebulae}

We now turn our attention to metal lines arising from the extended nebulae, which could further constrain the properties of the emitting gas (e.g. its metallicity or gas density). We focus on the two strongest transitions in our wavelength range, namely \CIV1549\AA, and \HeII1640\AA. Another useful transition is \OVI1034\AA, however this line is very close to the Ly$\beta$ and is in the quasar \lya\ forest, making the detection of a clean signal particularly challenging.

\begin{figure}
    \centering
    \includegraphics[width=0.50\textwidth]{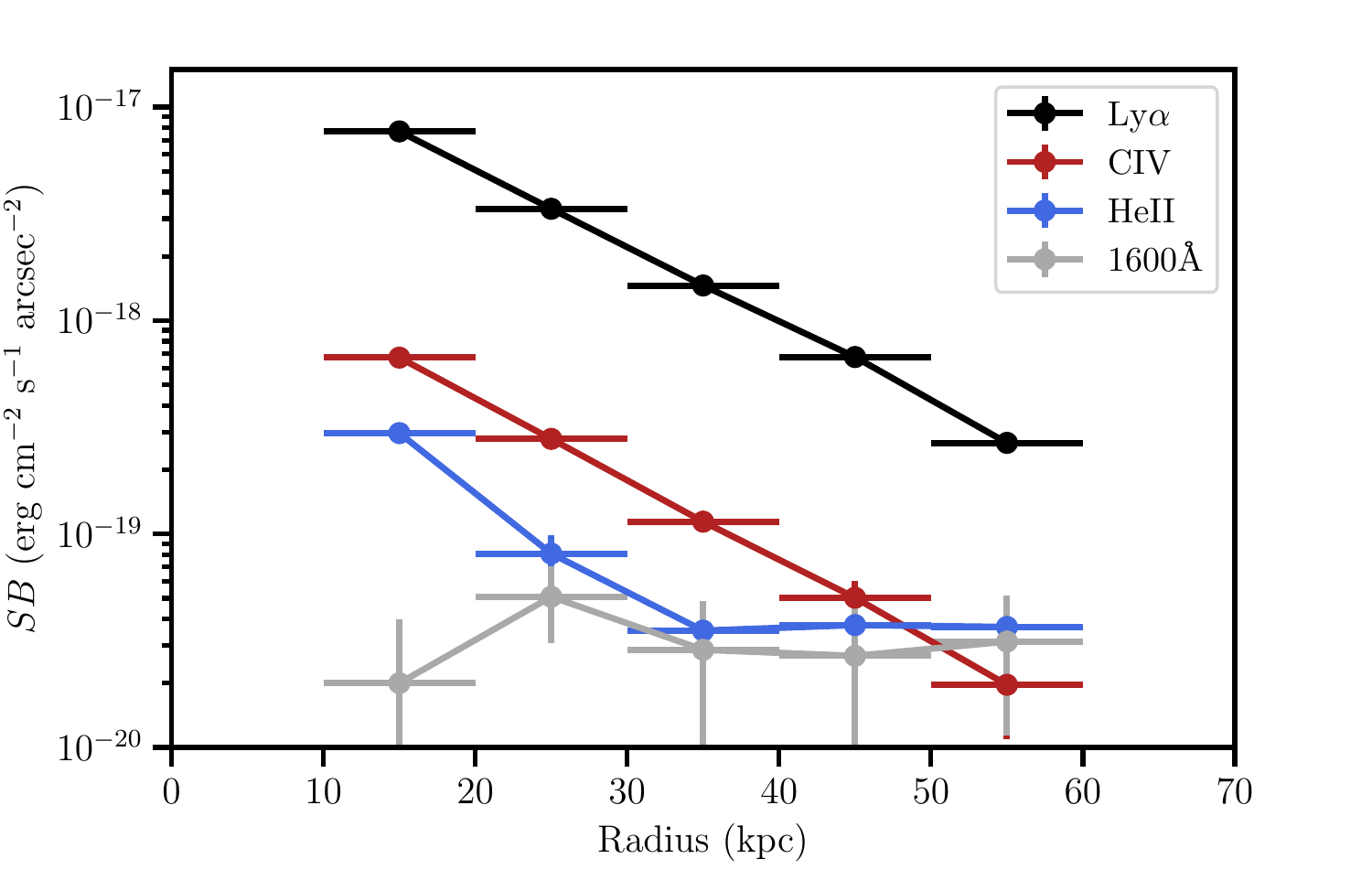}
    \caption{Average observed surface brightness profiles for the same transitions as in Figure \ref{fig:nebstack2D} obtained in circular annuli centered on the quasar position. The grey points from the stack at 1600\AA\ in the rest frame are an indication of the noise level at the wavelengths of \CIV\ and \HeII.}
    \label{fig:nebstackrad}
\end{figure}

The lines of interest are much fainter than \Lya, and we do not detect them in the spectra of individual nebulae. We therefore stack the signal from individual nebulae, thus boosting the final $S/N$. 
Our stacking procedure is applied identically to all transitions. First, we start by generating NB images with 30\AA\ width centered at the redshifted wavelength of each transition. We then convert the distance of each pixel from the quasar position into a comoving distance (in kpc), and we map the flux into a cosmological dimming corrected surface brightness (in \sbline). This choice of coordinates is driven by the large redshift range of our sample, where changes in surface brightness and radius as the Universe expands must be taken into account to optimally coadd the data. Then, we interpolate these images such that the quasar is positioned at the centre of a common grid. 

During the interpolation we rotate each field by 12 degrees around the quasar position to make sure that residuals of instrumental artefacts are randomly positioned in the final grid, thus optimally suppressing them. Lastly, we stack the positioned images using median statistics and we convert back to observed surface brightness and proper distances using the median redshift of the sample. 
\citet{Guo20} performed a similar stacking study of $\approx80$ nebulae from the \citet{Arrigoni-Battaia19} and \citet{Borisova16} samples. These authors found that the $S/N$ of extended emission in the stack is boosted if they stack only asymmetric nebulae along the \lya\ orientation. Our nebulae however are mostly symmetric and we did not find any $S/N$ improvement if we align the \lya\ emission before stacking. Therefore, we report the results of our random orientation stacking hereafter.

Figure \ref{fig:nebstack2D} shows the \Lya\, \CIV, and \HeII\ surface brightness maps of the two dimensional stacks, as well as a stack at 1600\AA\ in the rest frame to evaluate whether continuum subtraction residuals are present near the metal lines. We estimate the uncertainty in the profiles by bootstrap resampling of the full stacking procedure. In Figure \ref{fig:nebstackrad} we show the average radial surface brightness profiles for the same emission lines and the control stack. The inner 10 kpc of the profiles are dominated by the quasar PSF and are therefore masked both in the 2D maps and in the radial profiles. Besides a strong detection of extended \lya\ emission, we also detect \CIV\ extending up to $\approx$50 proper kpc from the quasars, with an integrated detection significance of $\approx 5\sigma$. An inspection of the profiles for each quasar did not reveal a significant detection of \CIV\ in any individual field. We also find a tentative detection of extended \HeII\ which reaches barely 2$\sigma$ significance.
To measure diagnostic line ratios in our stacks, we consider two regions at $10<R/{\rm kpc}<30$ and $30<R/{\rm kpc}<50$, where the inner limit is set by the quasar PSF and the outer limit is chosen to maximise the robustness of the line ratios, since at larger radii we only detect \lya\ emission. Our line ratios in the two radial bins are \CIV/\lya$= 0.073^{+0.015}_{-0.016}$ and $~0.092^{+0.025}_{-0.023}$, and  \HeII/\lya$= <0.031$ and $<0.032$. Uncertainties are from bootstrap resampling and upper limits are quoted at $1\sigma$. In Section \ref{sec:discussion_neb} we will discuss how this evidence constrains the physical conditions of the gas in these extended nebulae.

\subsection{The population of LAEs around the quasars} \label{sec:results_lae}

The combination of medium depth and large number of independent fields is a distinctive characteristic of the MAGG survey compared to most literature studies that examine either large samples of quite shallow exposures or deep observations of individual fields. In MAGG, we thus have an excellent dataset to study not only extended emission at the quasar redshift, but also the demographics and the properties of LAEs in the quasar environment.

\begin{figure}
    \centering
    \includegraphics[width=0.45\textwidth]{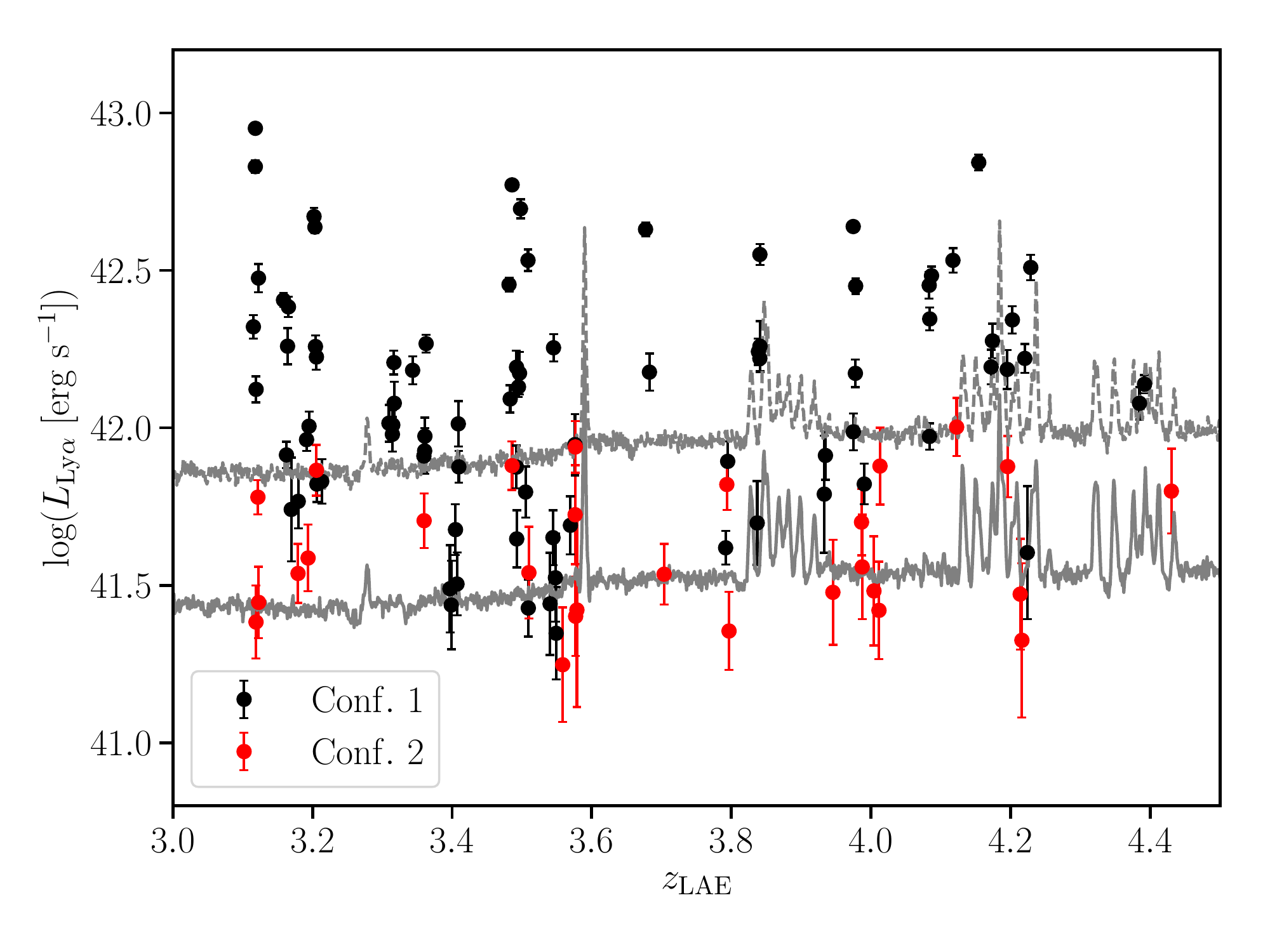}
    \caption{Luminosity-redshift distribution of the LAEs in the quasar environment from the 27 MAGG fields. Black and red points mark confidence 1 and 2 sources, respectively. The grey solid and dashed lines mark the 10\% and 90\% detection fractions, respectively. }
    \label{fig:lumvszlae}
\end{figure}

With the algorithms described in Section \ref{sec:detLAE} we built a final LAE catalogue including 113 sources. Their luminosity distribution as a function of redshift is shown in Figure~\ref{fig:lumvszlae}. Confidence 1 and 2 sources are shown as black and red points, respectively. Although our sample becomes more and more incomplete with decreasing LAE luminosity, Figure~\ref{fig:lumvszlae} shows that the majority of our LAEs are in the high-confidence class and have a luminosity above the 90\% completeness limit across the full redshift range, making our sample representative of LAEs in the quasar environment. Nonetheless, when we present statistical results we use the full sample (unless otherwise stated) corrected for incompleteness using the selection function from real source mocks (see Figure~\ref{fig:selfun}).

\begin{figure}
    \centering
    \includegraphics[width=0.45\textwidth]{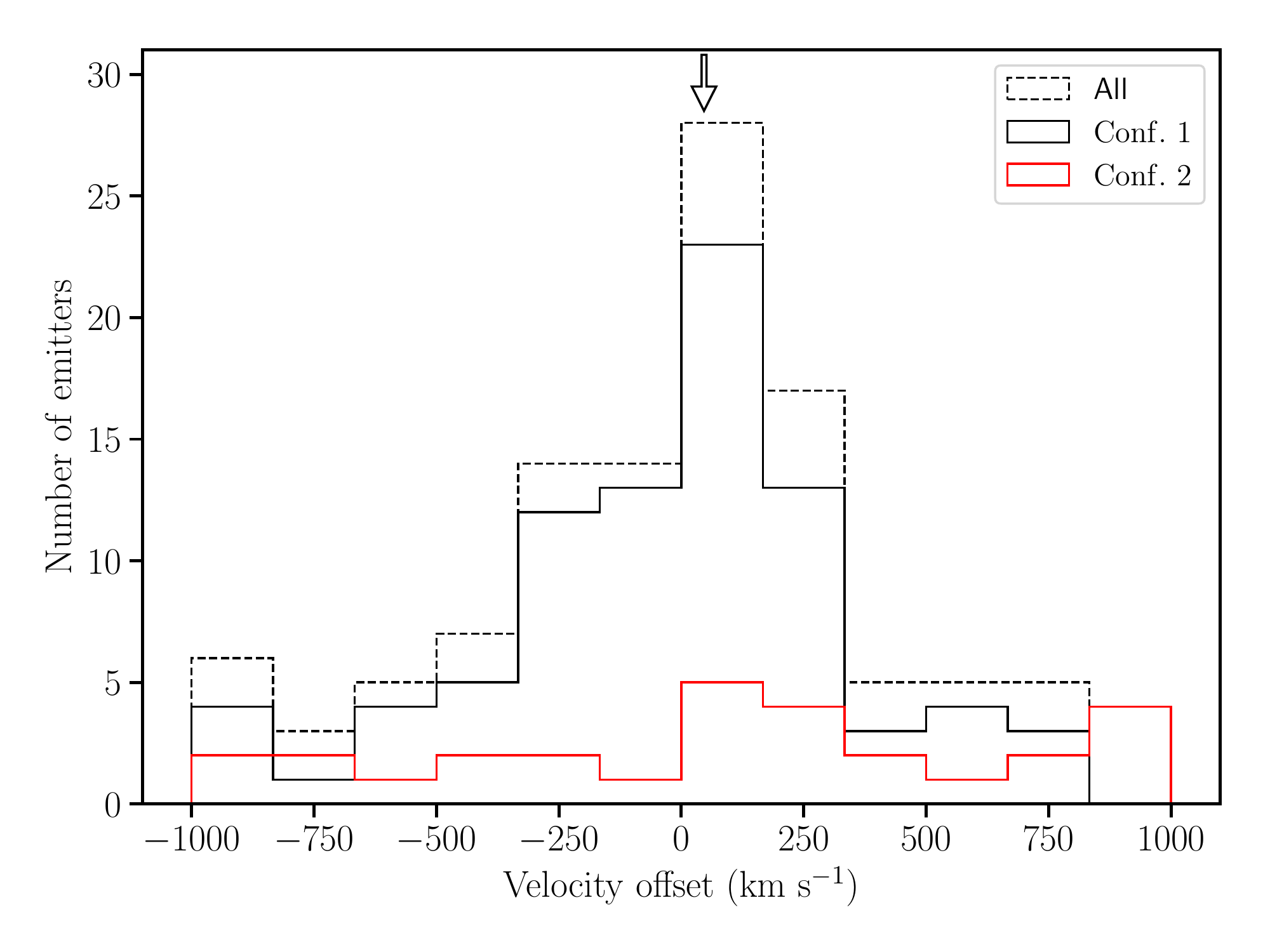}
    \caption{Velocity offset between the LAE redshift (from the \lya\ emission line) and the redshift of the extended \lya\ nebula in the same field. The dashed histogram is for the full sample, while the black and red solid lines are for confidence 1 and 2 sources, respectively. The empty black arrow marks the median offset for the full sample. }
    \label{fig:deltavLAE}
\end{figure}

In Figure \ref{fig:deltavLAE} we show the distribution of velocity offsets between the redshift of each LAE and the redshift of the nebulae in the same field. The distribution is peaked near zero with a median offset of $\approx 26$~\kms. It is known \citep[e.g.][]{Muzahid20} that the \lya\ line of LAEs is on average redshifted by $\approx 200$~\kms\ with respect to the systemic redshift. Despite the different physical conditions (density and ionization field) of the ionized gas in LAEs and in the extended nebulae, the fact that the average velocity offset between them is close to zero indicates that the \lya\ emission of the nebulae is affected by a similar amount of scattering and is redshifted with respect to the systemic redshift. 


Previous works have found that quasar hosts at $z>3$ are found in relatively massive halos with $M_h \approx 10^{12.5}{\rm M}_\odot$ \citep{Shen07,He18,Timlin18}, for which velocity dispersions of 200-300 \kms\ are expected. When focussing only on high confidence LAEs, we find that 86\% of them have a velocity offset within 300~\kms. \citet{Farina19} reported a median offset between the redshift of the extended \lya\ nebulae and the quasar systemic redshift of only 54 \kms. This result, combined with the fact that most of our LAEs are clustered at small velocity offsets with respect to the redshift of the extended nebulae, imply that the results presented here are insensitive to the choice of using the centroid of the extended nebulae emission as a reference instead of the quasar systemic redshift. It is therefore reasonable to assume that our LAE sample traces the near environment of quasars (see also Section~\ref{sec:LF}), where the quasar radiation could contribute to the ionization field. This population is not limited to true satellite galaxies within the virial radius of the host halo, but also includes galaxies beyond this radius which are nonetheless affected by the halo potential. The velocity distribution of low confidence (class 2) LAEs is flatter and broader with no distinct peak, possibly as a result of the larger uncertainty on individual redshifts due to the lower SNR.  We will further discuss the implications of the LAE clustering in velocity space on the environment of quasars in Section \ref{sec:discussion_lae}. 

\begin{figure}
    \centering
    \includegraphics[width=0.5\textwidth]{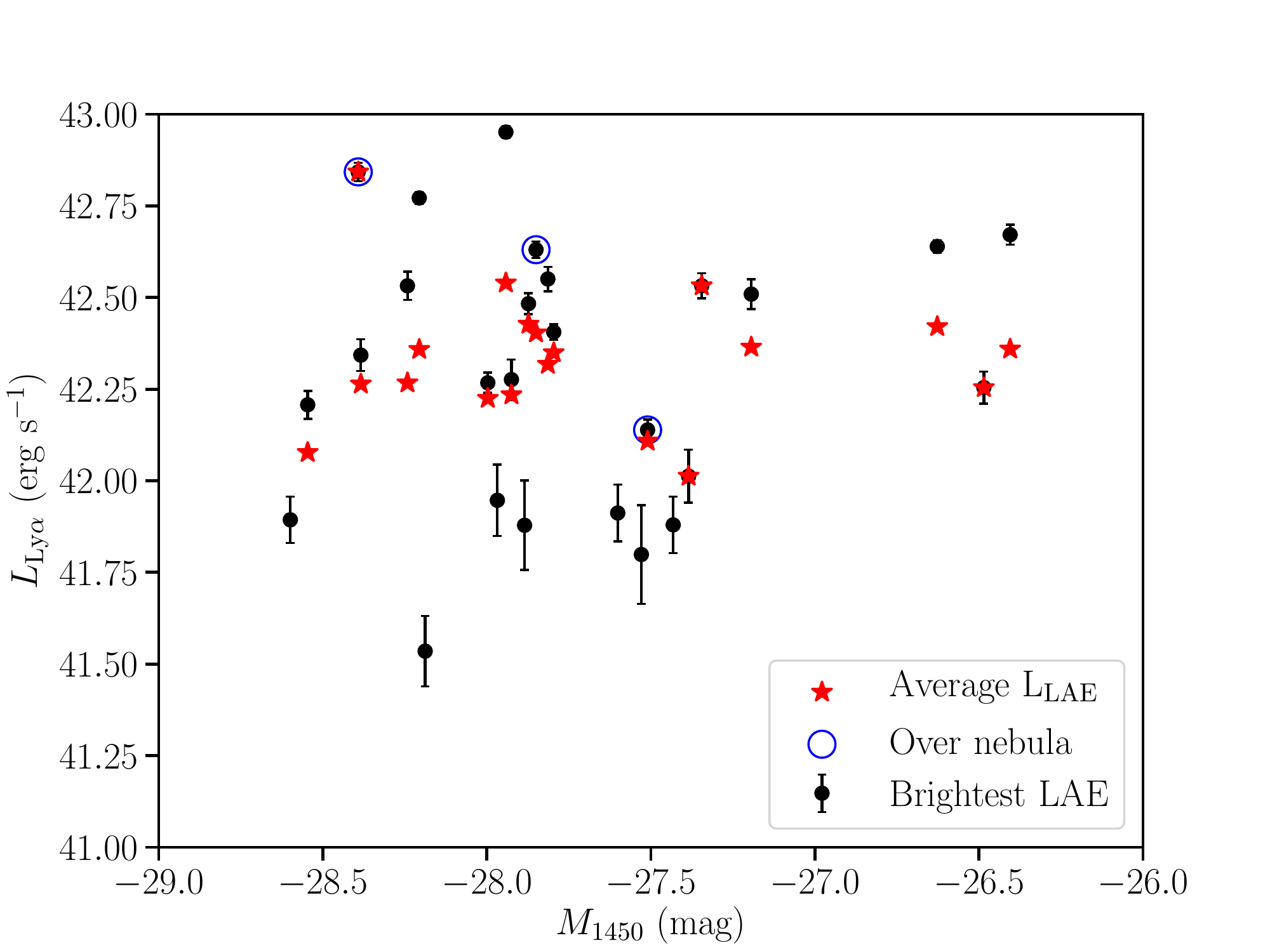}
    \caption{The luminosity of the brightest LAE (black points) in each quasar field as a function of the quasar absolute magnitude ($M_{\rm 1450}$). Emitters that are spatially overlapping with the \lya\ nebulae are highlighted by a blue circle. Red stars show the average LAE luminosity in each quasar field for LAEs brighter than $L_{\rm Ly\alpha}>10^{42.0}~\rm{erg~s^{-1}}$. This conservative threshold is chosen to make sure the selection function of LAEs is close to unity and therefore we do not miss a significant fraction of emitters in our extraction.}
    \label{fig:LAElumvs1450}
\end{figure}

We now investigate if the LAE luminosity correlates with the absolute magnitude of the quasars. We use two metrics to test if such a correlation exists. First, we plot in Figure \ref{fig:LAElumvs1450} the \lya\ luminosity of the brightest LAE in each field as a function of $M_{\rm 1450}$. The luminosity of the brightest emitter is thought to be a good indicator of the effects of the quasar ionization field on LAEs,
as the boosting induced by the quasar radiation on Ly$\alpha$ should lead to the presence of LAEs that are brighter then typically found in the field. Turning the argument around, the brightest LAEs are thus prime candidates for objects with boosted Ly$\alpha$. 
However, relying on just one LAE per field, this quantity is intrinsically noisy and subject to stochasticity. 

We therefore use a second metric designed to test the existence of a correlation between the average luminosity of LAEs in each field and the quasar absolute magnitude. We take the average \lya\ luminosity for emitters brighter than $L_{\rm Ly\alpha}>10^{42}~\rm{erg~s^{-1}}$, where our selection function shows that we have a highly complete LAE population across the full redshift range, and we show this quantity as red stars in Figure \ref{fig:LAElumvs1450}. A clear correlation between these two metrics of LAE luminosity and $M_{\rm 1450}$ is not found. 

We quantify the probability of the presence of a correlation (or its absence) by using the Spearman rank correlation coefficient. For the brightest LAE luminosity versus $M_{\rm 1450}$, the correlation coefficient is $\rho=0.13$, which turns into a probability that the two quantities are uncorrelated of $P(\rho |{\rm null}) = 0.51$. Similarly, when the average LAE luminosity is used, we obtain $\rho=0.08$ and $P(\rho |{\rm null}) = 0.73$, which confirms the lack of a significant correlation. We also find that the number of LAEs detected in each quasar field does not correlate with $M_{\rm 1450}$. In this case the correlation coefficient is $\rho=0.12$, with an associated $P(\rho |{\rm null}) = 0.54$.

This evidence could be interpreted as an indication that the luminosity of LAE population is not predominantly influenced by the ionization field of the quasars, as we will discuss in more detail in Section \ref{sec:discussion_lae}.  However, we also note that $M_{\rm 1450}$ is only a proxy for the flux of ionizing photons and the correlation with Ly$\alpha$ may be weakened by the scatter between $M_{\rm 1450}$  and the quasar ionizing luminosity, or by anisotropic effects in the quasar illumination field \citep{Hennawi07}.

\begin{figure}
    \centering
    \includegraphics[width=0.5\textwidth]{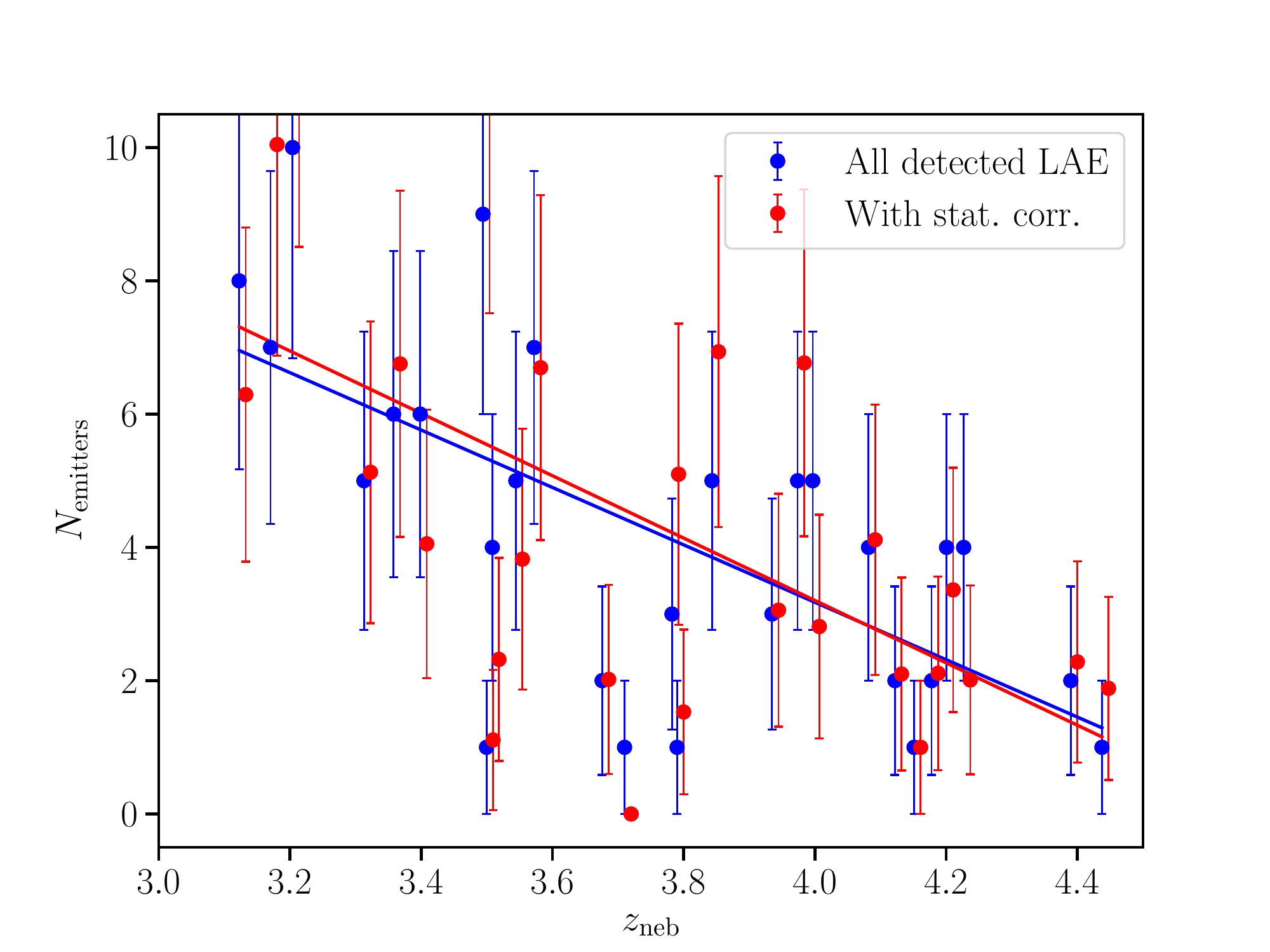}
    \caption{Number of LAEs in each quasar field as a function of redshift. Blue points are obtained using all detected LAEs, while red points are obtained using only LAEs with a detection probability $f_c(L,z)>0.20$ and a statistical correction for the sample incompleteness based on the selection function. The linear fits to these two samples show that the trend does not change if the statistical correction is applied.}
    \label{fig:LAEnumvsz}
\end{figure}

Next, we study the number of LAEs in each quasar field as a function of its redshift, as parametrized by the nebulae redshift in Figure \ref{fig:LAEnumvsz}. The blue points are obtained with the raw number of detected LAEs (both confidence 1 and 2) in each field and suffer from the sample incompleteness. To directly test the effects of incompleteness we make use of the  selection function derived in Section \ref{sec:selectionfunc}. We select only LAEs which have a probability of being detected above 20\%  ($f_c(L,z)>0.20$), to avoid being biased by a few sources with  very large weights when a statistical correction is applied. Because $f_c(L,z)$ is only weakly dependent on redshift, our threshold roughly correspond to a volume limited sample defined by $L_{\rm Ly\alpha}>10^{41.5}~\rm{erg~s^{-1}}$. We then sum the number of emitters corrected for their detection probability, in symbols $N_{\rm emitters} = \sum_i 1/f_c(L_i,z_i)$ where $i$ runs over the selected LAEs. In both cases the uncertainties are obtained from Poisson statistics, and we fit the data points with a linear function. 

From Figure \ref{fig:LAEnumvsz}, it is immediately clear that an increasing number of emitters are found with decreasing redshift of the quasar host. The trend is qualitatively the same if we consider the raw data or the sample corrected for the selection function (hereafter, we use the completeness corrected values). We also verified that this is not caused by a difference in the fraction of deblended emitters with redshift or objects very close to the extended nebulae. The interpretation of this trend is made difficult by the many effects which are simultaneously at play and that could create this signal (e.g. changes in surveyed volume or in LAE detectability due to quasar fluorescence).  We note however that the comoving volume we used to search for emitters around each quasar does not significantly change across the redshift range considered here and it is only 10\% bigger at $z=4.5$ compared to $z=3.0$. The different number of emitters found at the two extreme ends of the redshift range cannot be driven by a volume variation. Moreover, we also have not found a correlation between LAE luminosity or the number of LAEs in each field and quasar absolute magnitude.

The explanation for the higher number of LAEs in lower redshift fields can therefore reside in an evolution of the properties of LAEs. Unfortunately, at the depth of our observations, only 12\% of the LAEs are detected in continuum emission. \citet{Lofthouse20} calculated the $r-$band 90\% completeness limit of the MAGG survey to be 26.3 mag. If we model the star formation histories of LAE as single bursts of star formation occurring near the observed redshift \citep{Stark09}, using \citet{Bruzual03} stellar models and a \citet{Chabrier03} initial mass function we obtain that this survey limit corresponds to a stellar mass $M_*>10^{9}{\rm M_\odot}$. At fixed stellar mass, \citet{Schreiber15} shows that the average star formation rate of galaxies on the star forming main sequence increases by a factor of $\approx 2$ between $z=3.0$ and $z=4.2$. At such high redshift these results are based on more massive galaxies, however at lower redshift these authors have shown that the redshift evolution of the main sequence is mass independent. If our LAEs have the same mass across the redshift range we study, we should expect that there are more LAEs with a higher SFR, and therefore a higher \Lya\ luminosity, with increasing redshift. 

We are therefore left with the following hypothesis: that there are more LAEs above a given stellar mass in the environment of quasars as the Universe evolves. However, the mean \Lya\ luminosity of our LAE sample is constant at $10^{42}~\rm{erg~s^{-1}}$ if we split the redshift range in three equally spaced bins. This result requires LAEs to grow in mass or clustering without appreciably increasing their \Lya\ luminosity such that we see more of them around lower redshift quasars compared to higher redshift.
As previously noted, quasar-induced Ly$\alpha$ fluorescence further complicates this picture. A direct continuum detection and a robust estimate of the stellar mass of LAEs are required to quantitatively confirm this interpretation. 

\subsection{Spatial alignment of LAEs}

Another diagnostic made accessible by the MUSE integral field observations is the spatial distribution of LAEs around the central quasar host. We can then study two scenarios for the assembly of quasar haloes: that galaxies are preferentially aligned along cosmic filaments or that they are more spherically distributed in the halo. Indeed, when studying the environment of one strong hydrogen absorber in MAGG \citet{Lofthouse20}, found an alignment between the positions of the absorber and LAEs detected with MUSE (see also \citealt{Fumagalli16}, and \citealt{Mackenzie19}). 

\begin{figure}
    \centering
    \includegraphics[width=0.5\textwidth]{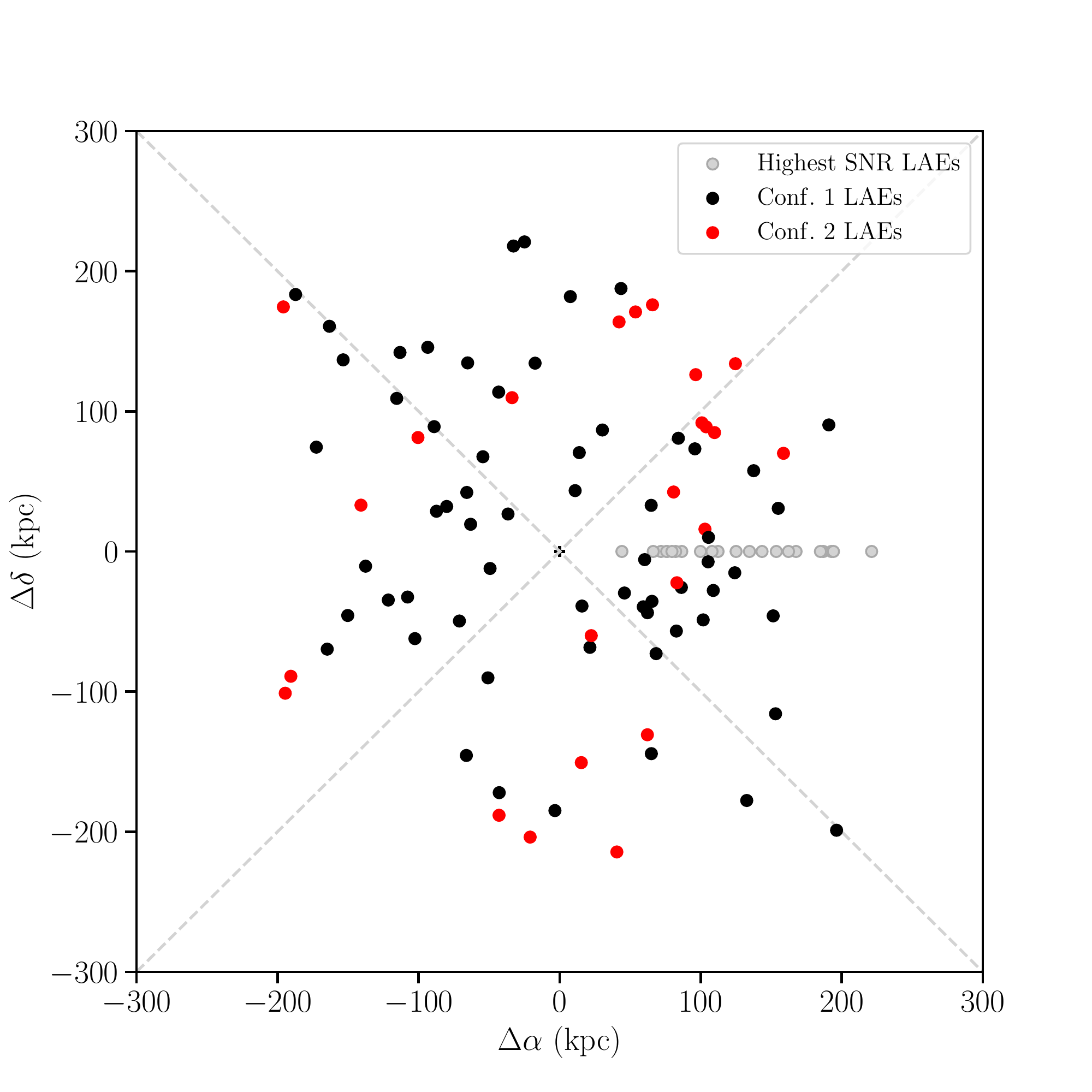}
    \caption{Distribution of LAEs spatial offsets from the quasar position. Each field has been rotated such that highest $S/N$ LAE in each field (grey points) appears along the positive x-axis. Black and red points identify confidence 1 and 2 sources, respectively. The grey dashed lines divide the space into four identical quadrants in which we evaluate the number density of LAEs.}
    \label{fig:align}
\end{figure}

Figure~\ref{fig:align} shows the positions of all the detected LAEs relative to the position of the central quasars. Individual sources have been positioned by means of a rotation around the quasar position such that the highest $S/N$ LAE in each field appears along the positive x-axis. If the LAEs are aligned in a filamentary structure this would result in an overdensity of points near the x-axis compared to the number of objects in the orthogonal direction. Here, we only plot systems with $\geq 2 $ sources in a single field. 

When considering all our sources we find 72 galaxies in the 45 deg cones straddling the x-axis compared to 36 in the vertical cones. However, when the highest $S/N$ emitters (which are forced to appear in the right hand cone) are removed, the number of LAEs aligned near the x-axis reduces to 48. Taken at face value, this number might indicate a mild excess of aligned emitters compared to a random distribution. The fraction does not change if we only consider confidence 1 sources. Indeed, only 12 LAEs with confidence 2 would be removed from both the spatial regions we consider. 

\begin{figure}
    \centering
    \includegraphics[width=0.5\textwidth]{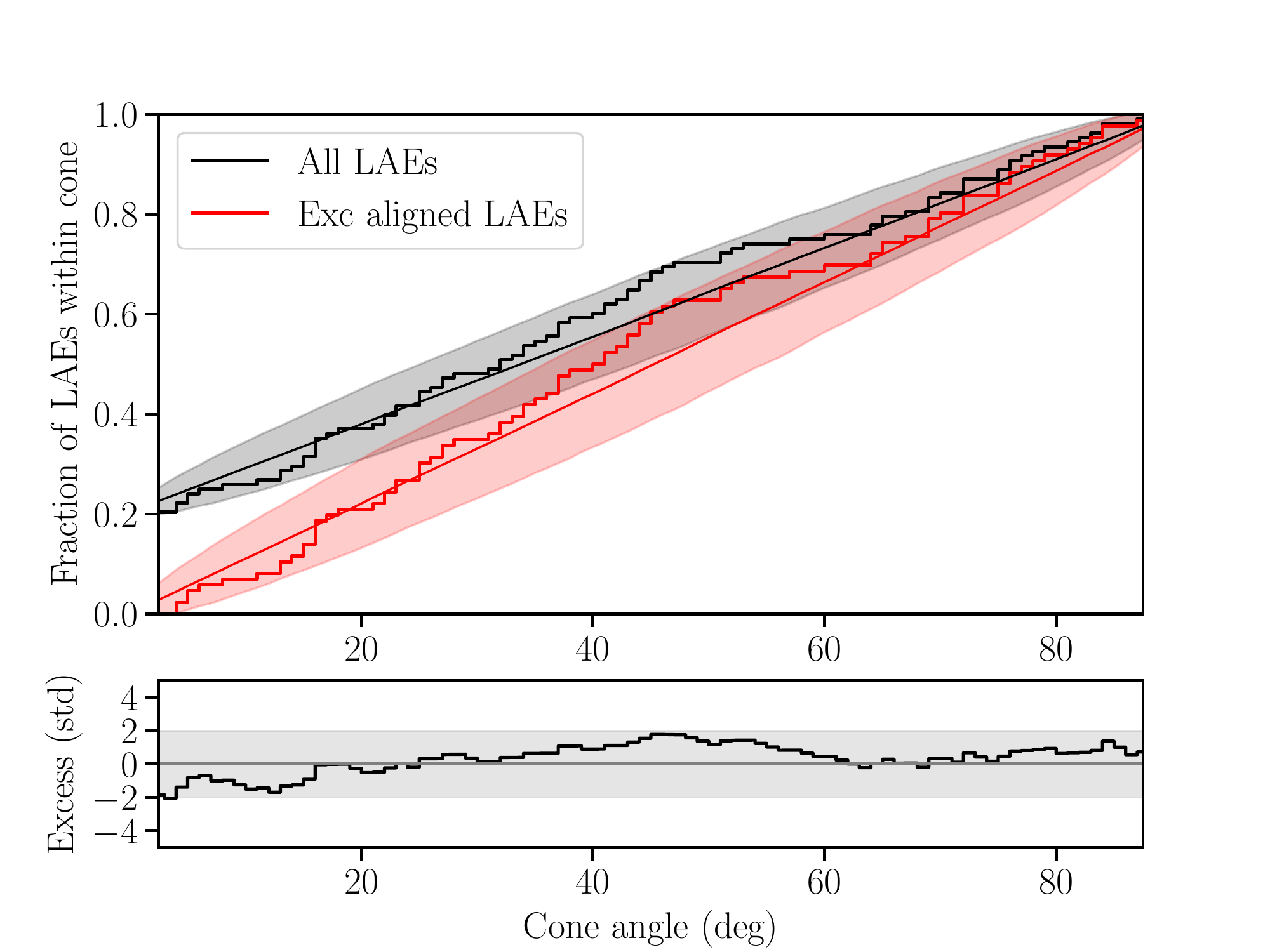}
    \caption{Top Panel: fraction of LAEs within the biconical region opening around the x-axis as a function of opening angle. The thick black solid line includes all LAEs, while the red line is obtained after excluding the highest $S/N$ LAE for each field, which is forced to appear along the positive x-axis. The thin solid lines and the shaded regions are respectively the median and the 2$\sigma$ confidence region obtained after randomizing the LAEs positions. Bottom panel: excess alignment of LAEs in units of standard deviations  compared to a random distribution.}
    \label{fig:fracalign}
\end{figure}

To study the alignment statistics in more detail, we evaluate the fraction of LAEs within the x-axis cone as a function of opening angle and we compare it to a mock sample obtained by randomising the LAE positions, therefore assuming no alignment. This diagnostic, as shown in the top panel of Figure \ref{fig:fracalign} indicates that an excess of aligned LAEs appears for opening angles between 40 and 60 deg. However this excess is not statistically significant above the 2$\sigma$ level when we consider 10,000 realisations of the random mock sample. As a result, we must conclude that there is no strong evidence of a preferential alignment of LAEs along filaments in the quasar environment, but rather that the emitters are almost randomly distributed around the central quasar as it would happen for galaxies orbiting or falling in a larger halo. 

\begin{figure}
    \centering
    \includegraphics[width=0.5\textwidth]{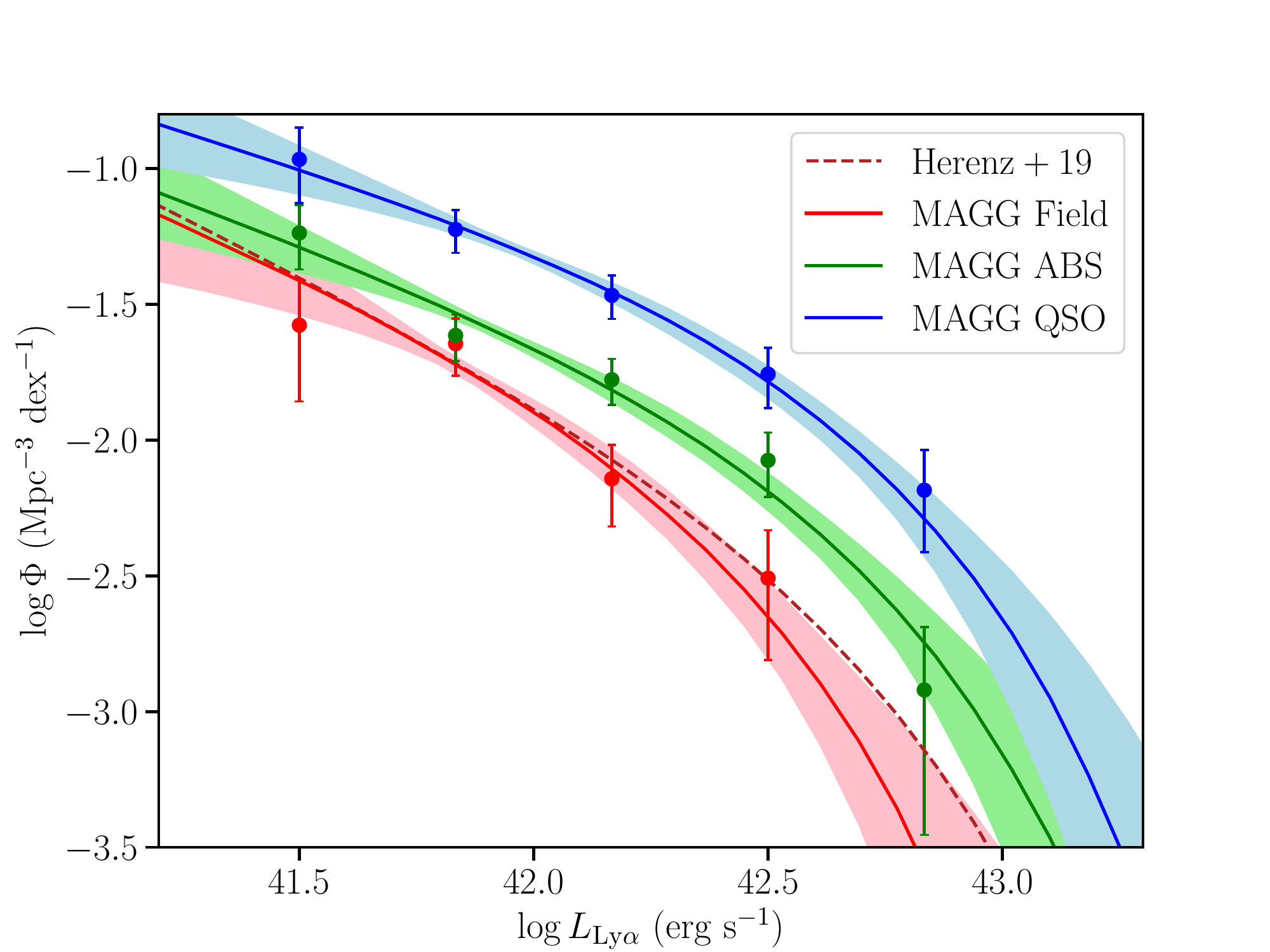}
    \caption{Differential LAE LF from the MAGG sample in the quasar environment (QSO, blue points and lines), near strong hydrogen absorbers (ABS, green points and lines), and in a control field sample (red points and lines). The non-parametric $1/V_{\rm max}$ binned estimate of the LF is given by the filled points with error bars, while the solid line and shaded areas mark the best fit and 1$\sigma$ confidence regions for the Schechter LF parametrisation. The red dashed line is the best Schechter fit of \citet{Herenz19}, who studied the field LAE LF at $2.9<z<6.7$.}
    \label{fig:lumfunc}
\end{figure}

\subsection{An enhancement in the number density of LAEs in the quasar environment} \label{sec:LF}

So far, we have shown that the spatial distribution and kinematics of LAEs can be interpreted with them being affected by the quasar halo potential. However, we have still not investigated how the LAE number density depends on the environment and if there is any evolution between the quasar environment and the general field. We can statistically characterise the population of LAEs by means of their luminosity function (LF). The differential luminosity function $\phi(L,z)$ counts the number of galaxies (in our case LAEs) per unit volume as a function of luminosity and redshift. Building upon the recent determination of the field LF by \citet{Herenz19} who did not find an evolution of the LAE LF in the redshift range $2.9<z<6$, and the fact that our sample of LAEs is relatively small, we assume the effects of redshift evolution to be negligible. Therefore, we build the LAE LF including all our quasar fields, i.e. with an effective redshift range $3.2<z<4.4$.

We use a non-parametric LF estimator, namely the $1/V_{\rm max}$ estimator as proposed by \citet{Schmidt68} and \citet{Felten76}, and further modified to account for a redshift- and luminosity-dependent selection function as it is the case with our data \citep{Fan01, Herenz19}.
To simplify the notation in what follows, we define the base 10 logarithmic luminosity as $\tilde L = \log_{10} L$. Within this formalism the differential LF can be approximated in bins of luminosity as follows:
\begin{equation}
    \phi(\langle \tilde L_{\rm Ly\alpha} \rangle) = \frac{1}{\Delta \tilde L_{\rm Ly\alpha}}\sum_i \frac{1}{V_{\rm max,i}}
\end{equation}
where $<\tilde L_{\rm Ly\alpha}>$ is the average \lya\ luminosity of a bin, $\Delta \tilde L_{\rm Ly\alpha}$ is the width of the bin, and the sum is over all sources $i$ in that bin. Here, $V_{\rm max,i}$ is the survey volume weighted by the selection function over which a given source can be detected \citep{Johnston11}. This is defined by
\begin{equation}
  V_{\rm max,i} = \omega \int_{z_{\rm{min}}}^{z_{\rm{max}}} \delta_{\rm obs.}(z) f_c(L_i,z)\frac{dV}{dz}dz\:,
\end{equation}
where $f_c(L_i,z)$ is the redshift-dependent selection function at the luminosity of source $i$, which we take from our simulations of real sources, and $dV/dz$ is the differential comoving volume element \citep{Hogg99}. Traditionally, $\omega$ is the angular area of the survey, in our case the search for LAEs has been carried out only in a $\pm 1000$~\kms\ window centered on $z_{\rm neb}$ for each field. Therefore the effective survey area is equal to a single MUSE field (0.97 arcmin$^2$) and $\delta_{\rm obs}(z)$ is a function that is equal to the number of fields where the emitters have been searched for at a given redshift (and therefore it is zero where no search has been done).
Lastly, the uncertainty for each bin is given by
\begin{equation}
    \sigma \left[\phi( \langle \tilde L_{\rm Ly\alpha} \rangle) \right] = \sqrt{\frac{1}{\Delta \tilde L^2_{\rm Ly\alpha}}\sum_i \frac{1}{V^2_{\rm max,i}}} \:.
\end{equation}
We use these estimators in bins of width = 0.25 dex in $\tilde L$ and we remove from the sample those LAEs with $f_c(L,z)<0.1$ because their large photometric uncertainties would result into largely uncertain $V_{\rm max}$ weights.

We show the non-parametric LF estimate of LAEs in the quasar environment as the blue points in Figure \ref{fig:lumfunc}. In order to compare to other surveys and other environments in the MAGG survey, we fit the LF assuming a parametric \citet{Schechter76} function: \begin{equation}
    \Phi(L) = {\rm ln}(10)\Phi^* 10^{(\tilde L-\tilde L^*)(1+\alpha)}{\rm exp}(-10^{(\tilde L-\tilde L^*)})
\end{equation}
This procedure requires no binning of the data. Following the formalism of \citet{Mehta15} and their original maximum likelihood estimator, the probability of detecting a galaxy with $\tilde L$, given the LF, is:
\begin{equation}
    P(\tilde L_i) = \frac{\Phi(\tilde L_i)V_{\rm max}(\tilde L_i)}{\int_{\tilde L} \Phi(\tilde L_i)V_{\rm max}(\tilde L_i)d\tilde L}
    \label{eq:ProbLF}
\end{equation}
 
The likelihood function for the full sample is the product of the individual probabilities, and the posterior distribution and best fit parameters have been obtained using the {\sc MultiNest} Bayesian algorithm \citep{Feroz08,Feroz13}. Because the probability of detecting a galaxy involves the ratio between the differential and integrated LFs, the normalization cannot be determined by the likelihood maximization procedure. However, given $L^*$ and $\alpha$, the value of $\Phi^*$ is uniquely determined by the ratio of the number of LAEs in the sample and the value of the denominator in Equation \ref{eq:ProbLF}. 

We show the differential LAE LF from our quasar environment sample (QSO sample) in Figure \ref{fig:lumfunc}. The non-parametric $1/V_{\rm max}$ binned estimate of the LF is given by the blue points with error bars, and the solid line and shaded areas mark the best fit and 1$\sigma$ confidence regions for the Schechter LF parametrisation.
We then repeat exactly the same procedure (including the method to identify emitters, to obtain total fluxes and redshifts and to build the LF) for two different environments in the MAGG data. First, we analyse the volume around the high column-density hydrogen absorbers including Lyman Limit systems (LLSs) and Damped \Lya\ absorbers (DLAs; ABS sample hereafter) from the study by Lofthouse et al. (in prep.). This sample includes 117 LAEs near 60 absorption line systems in the same redshift range of our QSO sample. We further search for LAEs in each MAGG field, using two 1000 \kms\ windows randomly placed at redshift $3.0<z<4.5$ to build a control sample (field sample hereafter). These redshift slices are chosen not to overlap with the QSO or the ABS samples and lead to the discovery of 52 LAEs in total. 

The resulting differential LFs are shown with green and red points and lines, respectively. We compare our field LF with the Schechter fit of \citet{Herenz19}, who studied the field LAE LF at $2.9<z<6.7$ with MUSE data (red line in Figure \ref{fig:lumfunc}), finding  excellent agreement. This validates our entire procedure of LAE identification and flux estimate, since the \citet{Herenz19} work has used different algorithms for the source identification and flux estimates. We note that, recently, \citet{Guo20a} presented the largest spectroscopic sample of LAE at a fixed redshift  $z=3.1$, however their LF does not extend significantly below $L^*$ and is thus difficult to compare with our deep MUSE data.

\begin{table}
    \centering
    \begin{tabular}{lccc}
    \hline
        Sample & $\alpha$ & $\log(L^*)$ & $\log(\Phi^*)$  \\
        & & $({\rm erg~s^{-1}})$ & $({\rm Mpc^{-3}~dex^{-1}})$ \\
        
    \hline
         QSO   & $-1.52\pm0.24$ & $42.74\pm 0.29$ & $-2.46\pm0.10$ \\
         ABS   & $-1.60\pm0.27$ & $42.73\pm 0.34$ & $-2.90\pm0.12$ \\
         Field & $-1.68\pm0.43$ & $42.50\pm 0.45$ & $-3.02\pm0.15$ \\
    \hline
    \end{tabular}
    \caption{Median marginalised Schechter function parameters for the \lya\ LF in the three environments studied in this work.}
    \label{tab:LFfit}
\end{table}

\begin{figure*}
    \centering
    \includegraphics[width=0.95\textwidth]{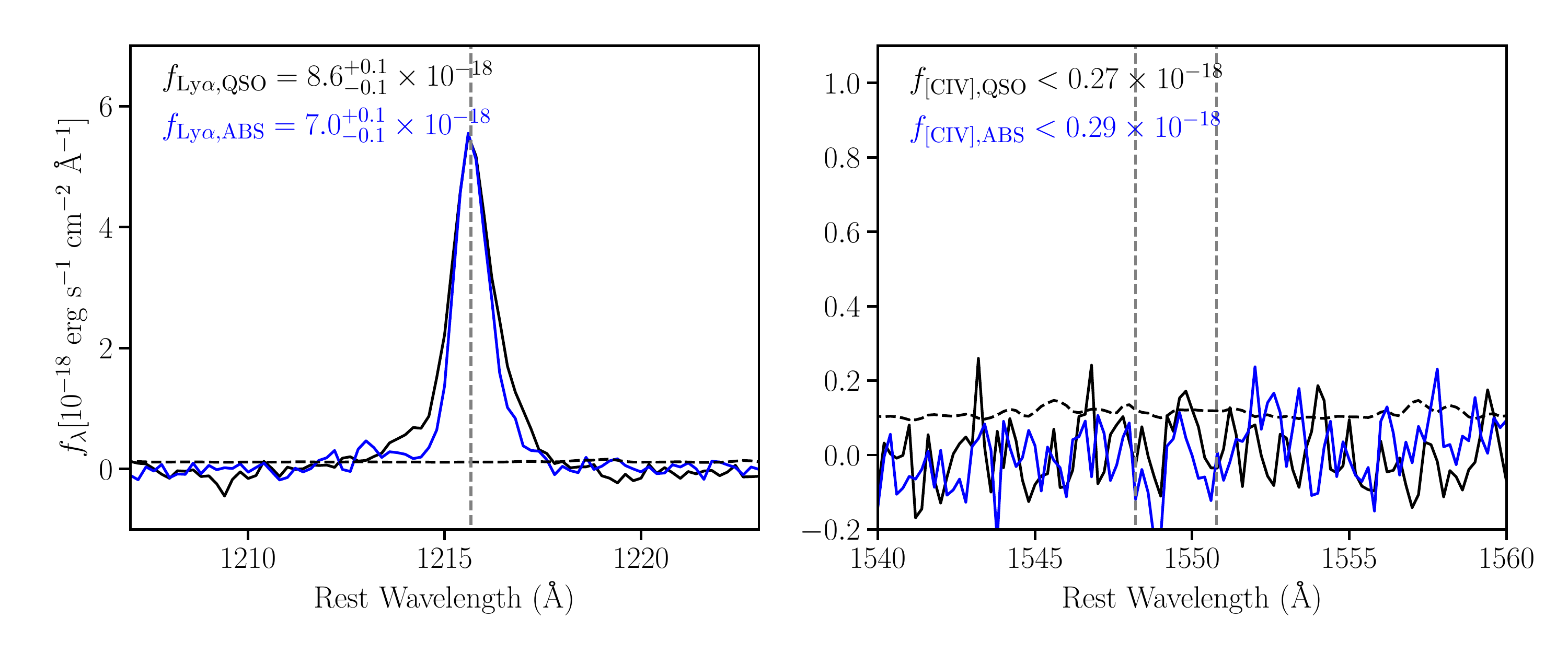}
    \caption{Median spectral stack for the LAEs detected in the QSO environment (black solid line) and in the vicinity of hydrogen absorbers (LLSs+DLAs, blue solid line). The spectra near the \Lya\ and the \CIV\ lines are shown in the left and right panels, respectively. Integrated fluxes are given in units of \flline. The grey dashed vertical lines show the rest-frame wavelength of the \Lya\ and the \CIV\ doublet lines. The black dashed line shows the spectral noise for the QSO stacked spectrum.}
    \label{fig:lae_specstack}
\end{figure*}

The median marginalised values for the MAGG Schechter function parameters are shown in Table \ref{tab:LFfit}, for the three environments we studied.
The values of $\alpha$ and $L^*$ are consistent across the different environments within their 1$\sigma$ uncertainties. However, the normalization is higher in the QSO sample by a factor 2.8 and 3.6 compared to the ABS sample and to the Field sample, respectively. This implies a larger number of LAEs near quasars with no significant difference in their \lya\ luminosity distribution. We also verified that the QSO LF fit parameters remain consistent with the full sample (within the uncertainties) both if we remove LAEs spatially overlapping with the extended nebulae and if we restrict to LAEs within 150 kpc from the quasars. \citet{Arrigoni-Battaia19} also studied the density of LAEs in the environment of quasars in the QSO MUSEUM survey. By comparing to the number densities expected from the field LF these authors found that the number of detected LAEs around quasars is consistent with the expectations from the field. Although this result appears in tension with the overdensity detected in our dataset, we argue that their shallower survey combined with a more conservative LAE extraction method and the completeness function estimates based on point sources (as opposed to the extended sources used in our work) could underestimate the number density of LAEs in the quasar environment.
  
A significant result of our analysis is the larger number of LAEs in the QSO samples (as well as the ABS sample, see Lofthouse et al. in prep) compared to the field. We can interpret the offsets between the LFs with a halo mass difference, indeed strong hydrogen absorbers are expected to live in more massive haloes compared to relatively isolated star forming galaxies \citep{Perez-Rafols18, Mackenzie19}, and the same argument applies to the bright quasars in the MAGG sample \citep{Shen07, He18, Timlin18}. However, the interpretation of the LF offset between the ABS and QSO environments is made more difficult by the concurrent possible roles of halo mass and the quasar ionization field which could boost the \Lya\ luminosity of LAEs in its vicinity.
In Sections~\ref{sec:spatspecCGMLAE} and \ref{sec:discussion_radfield} we attempt to identify if the quasar radiation field has a detectable effect on the properties of LAEs in its proximity zone.

\subsection{The spatial and spectral properties of the CGM of LAEs} \label{sec:spatspecCGMLAE}

We start by studying the spectral properties of LAEs in the QSO sample and we compare them to the ABS sample. We create rest-frame median stacked spectra of the LAEs in both samples using the spatial pixels from the {\sc CubEx} detection map. Figure \ref{fig:lae_specstack} shows the stacked spectra near the \lya\ (left panel) and \CIV\ (right panel) emission lines for the QSO  (black line) and the ABS (blue line) samples, respectively. We integrate the \lya\ emission in a window of 8\AA\ centered on its rest-frame wavelength. Conversely \CIV\ emission is not detected and we obtain a 2$\sigma$ upper limit on its flux by assuming that each component of the doublet has the \Lya\ spectral shape and estimating the flux using Bayesian fitting.

We find that the \lya\ flux in the QSO stack is higher than in the ABS stack but only by $\approx 20\%$. This result is indeed consistent with the evidence that the shape of the LFs in the two environments is the same, which results in a similar stacked flux. 
If we focus on \CIV, we do not find a detection in either sample, indicating that either the LAEs in the QSO environment are not significantly affected by the harder ionizing spectrum coming from the central quasar or that the metallicity of the ionized gas is too low for \CIV\ to be detected.

\begin{figure*}
    \centering
    \includegraphics[width=0.95\textwidth]{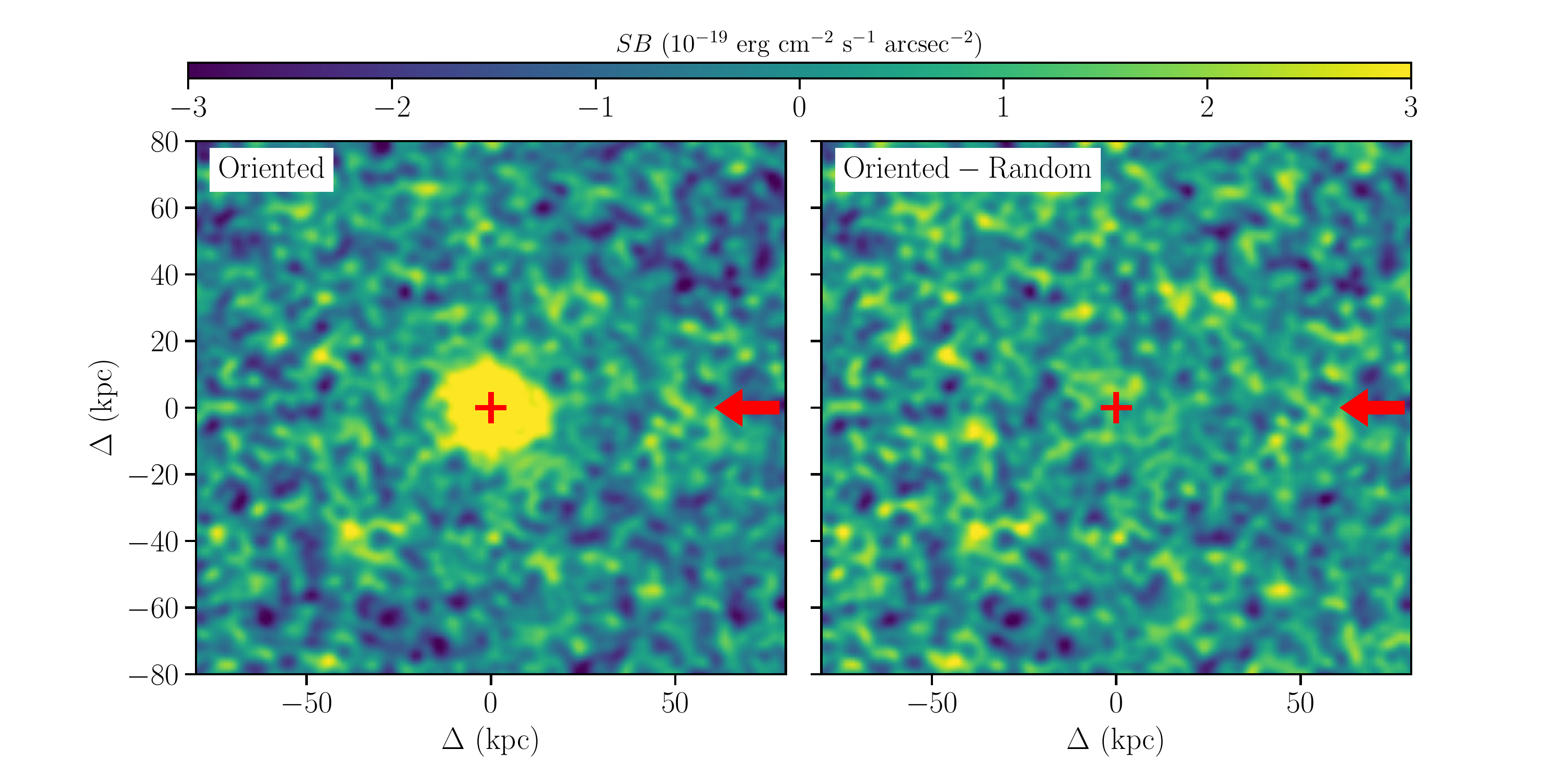}
    \caption{Left Panel: Median pseudo narrow-band \lya\ oriented stack for the LAEs detected in the quasar environment. Individual images have been oriented such that the quasar is always in the positive x-axis direction. The red arrow indicates the direction of incoming photons from the quasar, while the red cross is the origin of the coordinate system. Right Panel: the oriented stack after  subtraction of a combination of randomly oriented stacks to highlight the asymmetric features only. No clear asymmetry of the \lya\ emission is seen towards the quasar direction. }
    \label{fig:lae_nbstack}
\end{figure*}

Another property we can study for LAEs is their morphology. Thanks to the integral field data we can extract deep pseudo narrow-band images of the \lya\ emission to study if the CGM of LAEs in the QSO sample is affected by the quasar radiation. Indeed, if the Lyman continuum photons from the quasar illuminate the CGM of LAEs we could expect to see a faint and asymmetric (flourescent) \lya\ emission along the direction connecting each LAE with the quasar in their field. 

To reach very low surface brightness (SB) levels we adopt the stacking procedure developed by \citet{Gallego18}. In brief, for each LAE, we generate a pseudo narrow-band image centered at the \lya\ observed wavelength and $\pm 15\AA$ wide. We then resample the image such that the \lya\ emission of the LAE is at the center of a common grid and the quasar is aligned towards the positive direction of the x-axis. Similarly to the stacking of the extended nebulae, we perform this transformation in comoving spatial and surface-brightness coordinates, and we combine the data with median stacking. As a last step we return to proper distances and observed surface brightness at the median redshift of our quasar sample. This product, which we call the ``oriented' stack, is shown in the left panel of Figure \ref{fig:lae_nbstack}. 

The emission in this image is clearly dominated by a circularly symmetric component which needs to be removed if we want to study the possible presence of an asymmetric morphology. To do so, we follow \citet{Gallego18} and we generate a ``random'' stack obtained by averaging 200 stacks, each obtained with our LAE sample but with random orientations of the individual narrow-band images during the resampling procedure. We then subtract the random stack from the oriented one obtaining the map shown in the right panel of Figure \ref{fig:lae_nbstack}. The $2\sigma$ surface brightness limit of the residual image (calculated outside the region dominated by the LAE signal) is $3.2\times 10^{-19}$ \sbline and we do not detect any clear asymmetry in the direction of the oriented quasars above this value. We tested that stacking only galaxies within 150 kpc from the QSOs reduces the sample to 61 objects without yielding a detection of asymmetric signal. An additional caveat comes from the assumption that the \lya\ emission peak is not intrinsically shifted from the continuum peak. If this is not the case, our choice of centering the LAEs on the \lya\ peak could wash out an asymmetric signal. However, we verified that for the small fraction of continuum detected LAEs the continuum peak is within 0.3~arcsec from the \lya\ peak. 

Lastly we compared the circularly averaged radial profiles of the stack of LAEs in the QSO and ABS samples finding excellent agreement between them. We will leave a complete analysis of the shape of these profiles and including the modelling of the PSF effects to a dedicated future paper of this series. Overall, our results suggests that the CGM of LAEs in the environment of quasars does not appear to be asymmetric or more extended due to fluorescence from the quasar radiation, further reinforcing our previous findings based on the stacked LAE spectra.

\section{Discussion} \label{sec:discussion}
We have seen in Section \ref{sec:results_neb} that metal lines can be detected in extended nebulae around quasars, and in Section \ref{sec:results_lae} that the quasar haloes are hosting a rich population of LAEs. We now turn to the physical interpretation of these results using models for both the physical conditions of the extended nebulae around the quasars and the evolution of the LAE LF across different environments.

\subsection{Line ratio diagnostics in extended nebulae} \label{sec:discussion_neb}

The CGM of quasars is most likely multiphase and has a complex density structure. This complexity poses an obvious challenge to modelling, requiring for instance a sophisticated set of radiative transfer calculations in realistic high-resolution hydrodynamic models \citep[see e.g.][]{Buie20}.  
This approach is not currently readily available especially for the quasar CGM, and modelling efforts usually rely on approximated photoionization modelling in a single phase or clumpy medium. Indeed, following this strategy, line ratios have been widely used to estimate the metallicity and ionization conditions of gas in high-redshift radio galaxies, AGNs, and star-forming galaxies \citep{Nagao06,Nagao06a,Dors14,Dors19,Arrigoni-Battaia15,Matsuoka18,Nakajima18,Cantalupo19,Guo20,Travascio20}. We now discuss what we can learn from the \Lya, \CIV, and \HeII\ emission from the extended nebulae in terms of the metallicity and density of the CGM gas.

We first concentrate on the \HeII/\lya\ emission line ratio. \citet{Cantalupo19} studied this line ratio in the Slug nebula finding a value of $\approx 8\%$ and an upper limit of $\approx 1\%$ at physical distances that have been derived to be $\approx 270~$kpc and $\approx 900~$kpc from the quasar, respectively. These authors also interpreted the low observed \HeII/\lya\ line ratio as an indication that the emitting gas density is not constant, i.e. not a delta function as typically assumed. \citet{Cantalupo19} therefore concluded that the Slug nebula
could be composed by multiple structures at distances up to a physical Mpc along the line of sight. On small scales, the gas density distribution should also be clumpy as expected for example in a turbulent medium. In our stack of nebulae from MAGG quasars we find a 3$\sigma$ upper limit of 9\% for the \HeII/\lya\ ratio, which is as low as the values detected in the Slug nebula. If we follow the approach presented by \citet{Cantalupo19}, we can interpret this ratio in two non-mutually exclusive scenarios. First, the ionized gas might be at physical distances from the quasar that are much larger than the projected distances, which in turn would imply that the quasar ionization happens preferentially along the line of sight. Second, the CGM consists of gas with a very large range of densities, i.e. it is a clumpy and turbulent medium thereby confirming the results based on the Slug nebula with a larger statistical sample. It should be noted that, contrary to the MAGG nebulae, the Slug nebula is not detected in \CIV, and that the \HeII\ emission is associated to a compact off-centered source, therefore not symmetrically distributed around the quasar as in the MAGG tentative detection. The above analysis relies also on the detection of H$\alpha$ in the Slug nebula, which is useful to constrain the minor role of scattering in the Ly$\alpha$ line. However, lacking H$\alpha$ observations for the MAGG sample, we cannot definitely rule out scattering. In addition, Figure \ref{fig:neblyaprofiles}
shows a clear evolution between the \lya\ profiles of the CGM of $z\sim 2$ and $z>3$ QSOs, and there is evidence suggesting that the largest nebulae known are associated to extreme galaxy over-densities or the presence of strong radiation from active galactic nuclei \citep{Hennawi15, Arrigoni-Battaia18,Umehata19}. We cannot therefore exclude that the ionization conditions in the Slug nebula could be different from those in our MAGG sample. Thus, we caution that the above inference should be confirmed with further observations.

We next examine whether, following similar arguments, we can place constraints on the gas metallicity. 
We start by noting that our observed line ratios are consistent with those from
\citet{Guo20}, who have recently studied the emission lines in a stacked sample of extended quasar nebulae at $z\sim3$, and with those from \citet{Travascio20}, who mapped the \CIV\ emission in a single quasar nebula. \citet{Guo20} have detected \CIV, \HeII, and \CIII\ transitions, and have used them to estimate the gas metallicity in combination with simple photoionization models that do not take into account all the caveats discussed in more detail below. 
With this approach, they suggest that the CGM gas at $30-50$~kpc from the quasars has an average metallicity of $\sim 0.5~Z_\odot$ or higher. The presence of a metal-enriched gas phase is consistent with the direct measurements of the quasar CGM metallicity that have been obtained in absorption in several papers of the QPQ survey \citep{Prochaska13, Prochaska14, Lau16}, where an average metallicity between $1/10$ and $1/2$ solar is found. Moreover, the detection of metal lines at $z\approx3$ was predicted in post-processing also by cosmological simulations, e.g. in  \citet{Van-de-Voort13}. This work showed that the SB of \CIV\ is within the detection limits of our stack if the gas is moderately metal enriched ($Z>0.1 \rm{Z_\odot)}$ and the mass of the host halo is $M_{\rm halo} \gtrsim 10^{12} {\rm M_\odot}$, conditions that we could expect broadly for the bright quasars in our sample. 

Our observed line ratios would therefore suggest enriched gas in line with \citet{Guo20} if we were to adopt a similar modelling approach, e.g. using the models of \citet{Arrigoni-Battaia15} tailored to our quasar sample.
We verify, however, that these models are not able to reproduce the observed values of \lya, \CIV\ and \HeII\ simultaneously when assuming the gas at a distance similar to the projected distance, unless the metallicity is much greater than solar. More reasonable metallicities, in the range of the aforementioned previous works, are obtained if we include in input the line emissions from the central quasar (\lya\ and \CIV)  and we allow for resonant scattering of the \lya\ and \CIV\ photons. In this scenario, the resonant scattering could be an additional explanation to why the MAGG sample shows nebulae more extended in resonant lines than in non-resonant tracers (\HeII). These models are however approximations of the reality and complete radiative transfer calculations are required to precisely determine the balance between different powering mechanisms and therefore a value of metallicity.
In particular, a complete knowledge of the physical conditions of the CGM gas in extended nebulae (including its geometry and kinematics), as well as the detailed physics of ionized carbon are needed. 
Moreover, under the assumption that at least part of the line emission comes from collisional excitation, this emission would be very sensitive to the temperature. Many effects contribute to the temperature structure of the gas in the nebulae. While some of these are usually included in modelling efforts (e.g. photoionization and collisional excitation) others are more difficult to model, including the hydrodynamic interaction between inflows and outflows with the surrounding hot halo gas (which contribute to turbulence and shocks in the medium), and that are likely to depend on the halo mass of the quasar host and its gas cycle history.

Mindful of these strong caveats, it becomes very difficult to obtain a precise estimate of the metallicity in extended nebulae, which would depend on  many unverified assumptions and simplifications. 
Most notably, the role of scattering in powering the emission of the nebulae remains highly debated, and could in fact be an important component in the modelling. We therefore limit our conclusion to the fact that the extended gas is metal polluted, indicating some degree of enrichment of the medium for which we believe it is not currently possible to provide a specific value of metallicity.

\subsection{The galaxy environment of high-redshift quasars} \label{sec:discussion_lae}

\begin{figure}
    \centering
    \includegraphics[width=0.5\textwidth]{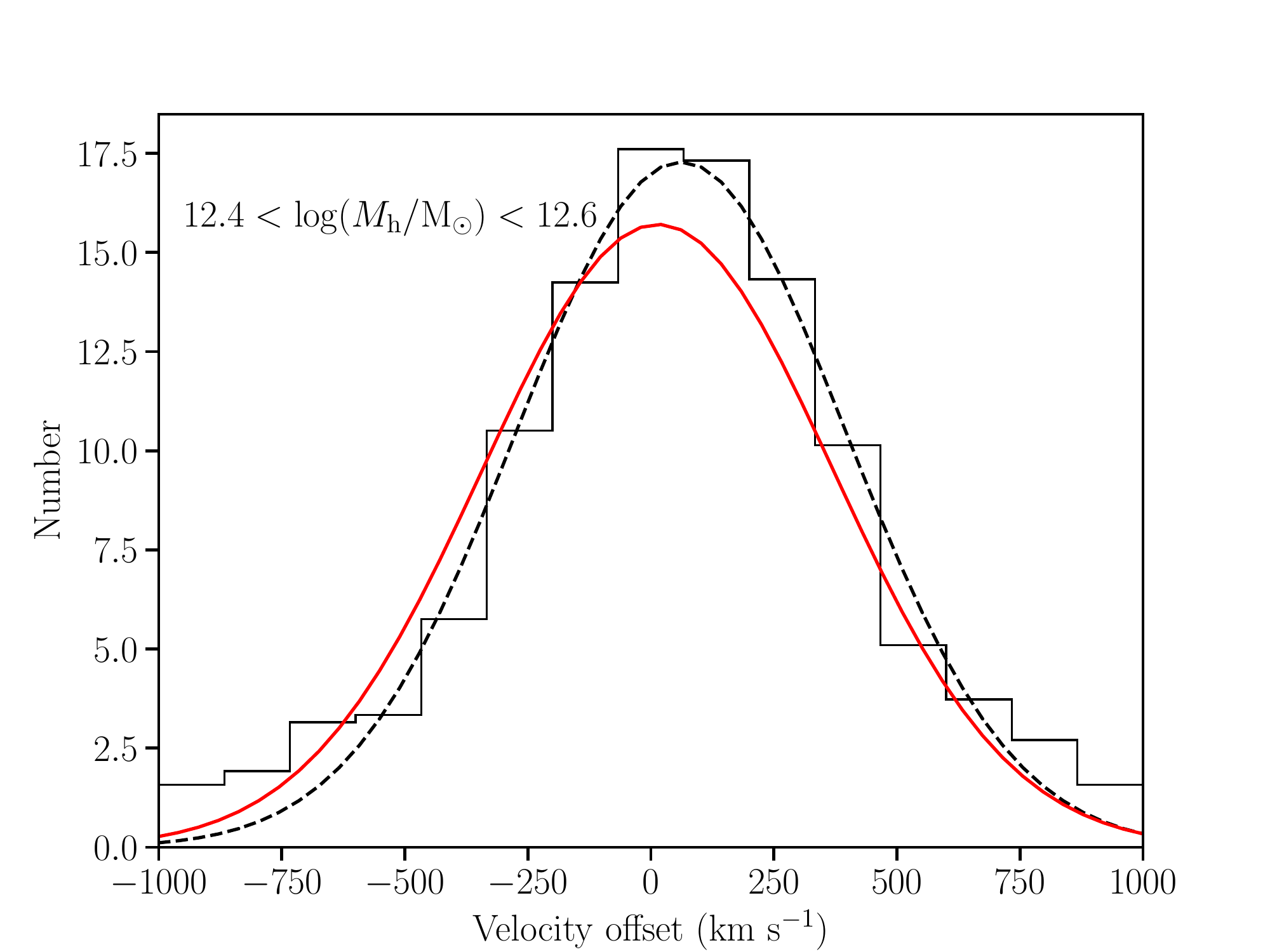}
    \caption{Velocity offset of a sample of mock LAEs (black histogram) selected from a semi-analytic model of galaxy formation (see text for the details of the selection). The histogram counts have been scaled to the number of detected LAEs in our MAGG quasar sample. The black dashed and red solid lines are Gaussian fits of the velocity distributions of the mock and the observed LAE QSO samples, respectively.}
    \label{fig:voffSAM}
\end{figure}

We now discuss in more detail the galaxy environment of $z=3-4.5$ quasars, starting from the results presented in Section \ref{sec:results_lae}, where we found a population LAEs likely orbiting in the quasar haloes. To confirm this interpretation and to evaluate the role of peculiar versus Hubble flow velocities in shaping the observed signal, we resort to the following simple modelling based on the the light-cones from the \citet{Henriques15} semi-analytic model of galaxy formation.

Starting with the common result that quasars are typically found in haloes of mass $M_{\rm h,QSO} \approx 10^{12.5}{\rm M}_\odot$ \citep{Shen07,Eftekharzadeh15,He18,Timlin18}, 
we select haloes in the range $10^{12.4-12.6}{\rm M}_\odot$ at $3.6<z<3.8$, corresponding to the average redshift of our quasar sample
The central galaxies of these haloes constitute our mock sample of quasar hosts. For our mock LAE sample, we further select all the galaxies above a stellar mass of $10^9~{\rm M}_\odot$ which are within a MUSE FoV and have a velocity offset (including redshift space distortions) within 1000 \kms\ from the quasar host. 

The distribution of velocity offsets, scaled to the number of LAEs in our sample is shown as the black histogram in Figure \ref{fig:voffSAM}. Among the neighbouring galaxies we find that 51\%, 73\% and 90\% of them are within 1, 2, and 3 times the halo virial radius, respectively. The first of these radii encloses what are typically called satellite galaxies of the quasar halo. We thus infer that the vast majority of our LAE sample is therefore within 3 virial radii of the quasar haloes, providing evidence that they are at least in infalling regions of the quasar environment. This is further confirmed by the fact that the full sample of selected galaxies has the same line-of-sight velocity distribution of satellites within the virial radius, indicating that the main halo potential dominates the kinematic of the LAE sample. Indeed, when we fit the velocity distribution with a Gaussian profile (black dashed line) and we overplot the Gaussian fit of the observed velocity distribution of LAEs shown in Figure \ref{fig:deltavLAE}, we find good agreement between the two Gaussian profiles. However, we need to take into account that part of the velocity dispersion estimate likely arise from the scattering of \lya\ photons both in the extended nebulae and in individual LAEs. As a result our observations appear indeed consistent with a typical halo mass $M_{\rm h,QSO} \approx 10^{12.5}{\rm M}_\odot$ as previously suggested in the literature, although the exact value might be lower given the above caveats on the observed velocity distribution.

Another estimate of the typical halo mass of our quasar sample can be obtained by comparing the number density of LAEs to the one presented by \citet{Trainor12}. To this end, we restrict to the LAEs that have a continuum detection and an $r-$band magnitude $<25.5$~mag to select a sample which is consistent with the one used by \citet{Trainor12}. We find a number density $\phi = (2.1-4.3)\times 10^{-3}~{\rm Mpc^{-3}}$, which compares to $\phi = (1.6-4.2)\times 10^{-3}~{\rm Mpc^{-3}}$ from \citet{Trainor12} after taking into account that their overdensity within 1 comoving Mpc in the transverse direction (the area covered by our MUSE observations) is a factor of two higher than in their entire FoV. These authors find that this number density, at $z\sim 2.7$ corresponds to an halo mass $M_{\rm h,QSO} = 10^{12.3\pm0.5}{\rm M}_\odot$ from a clustering analysis. While the number densities are in excellent agreement, it is possible that the typical halo mass of our sample is slightly lower due to the higher average redshift of our sample.

\begin{figure}
    \centering
    \includegraphics[width=0.5\textwidth]{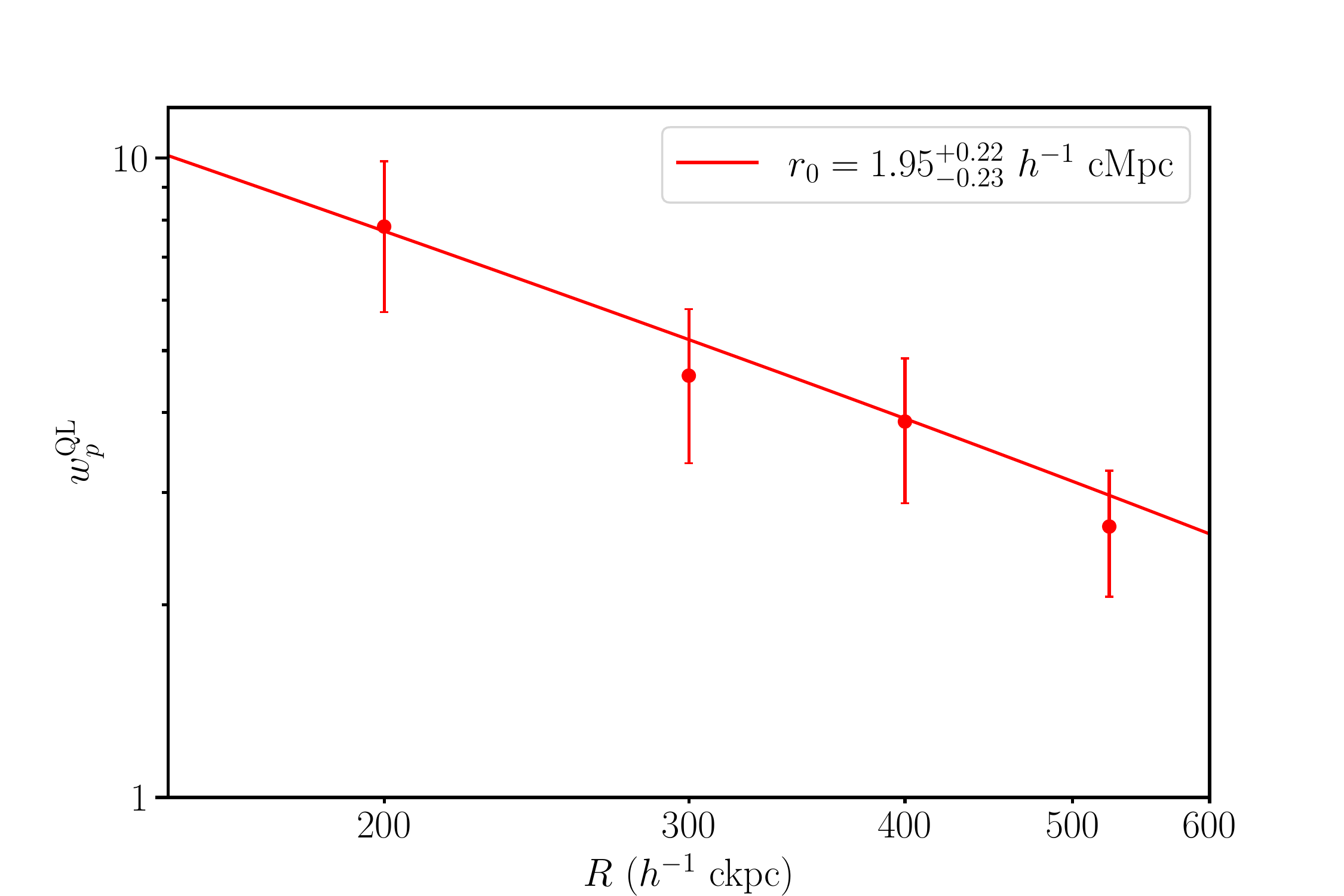}
    \caption{Quasar-LAE (QL) angular cross-correlation function. The uncertainties on individual datapoints are from Poisson statistics. The red line is a fit with the model in Equation \ref{eqXcorr} and a fixed slope  $\gamma=1.8$. The uncertainties on $r_0$ are obtained from bootstrap resampling with repetition of the individual quasar fields.}
    \label{fig:Xcorrfunc}
\end{figure}

The significant overdensity of LAEs in the quasar environment motivates us to compute the quasar-LAE cross correlation function, an additional estimate of the clustering properties of galaxies near quasars in our sample. The cross correlation function is defined as the excess probability of finding a galaxy in a volume $dV$ at distance $r$ from a quasar, compared to the probability of finding a galaxy of the same population in an average place of the Universe. The three dimensional correlation function usually assumes a power-law form $\xi = (r/r_0^{\rm QL})^\gamma$, where the QL superscript indicates the quasar-LAE cross correlation. As detailed in \citet{Trainor12}, the three dimensional correlation function is not directly measurable since line of sight velocities are affected by peculiar motions, a caveat that is very significant in our sample where most of the detected LAEs are affected by the quasar host halo potential. It is possible however to compute the reduced angular correlation function in a restricted velocity window around the quasar redshift. In this case the correlation function takes the form:
\begin{equation}
    w_p^{\rm QL}(R) = (r_0^{\rm QL} / R)^\gamma \times {}_{2}F_1 (1/2,\gamma/2,3/2,-z_w^2/R^2)
    \label{eqXcorr}
\end{equation}
where $R$ is the projected comoving separation between the quasars and the LAEs, ${}_{2}F_1$ is the Gaussian hypergeometric function and $z_w = (1000~{\rm km~s^{-1}})H_0^{-1}(1+z)^{-1}$ is the half-width of the redshift window in physical units. 

We estimate the angular correlation function in circular annuli centered on the quasar positions. We exclude the inner 150 $h^{-1}$ ckpc where the data is contaminated by the quasar PSF and we make sure that the largest annulus is entirely within the MUSE field of view across the full redshift range of our sample. Lastly, we use the field LAE sample to evaluate the average number of LAEs expected in each annulus. The quasar-LAE cross correlation function is shown in Figure~\ref{fig:Xcorrfunc}, where the red line is a fit of the data with Equation~\ref{eqXcorr}. Given the small dynamic range in projected radius we fix the slope of the correlation function to $\gamma=1.8$ as typically done in similar studies \citep[e.g.][]{Diener17,Garcia-Vergara19}. We find a positive correlation signal, confirming the presence of a significant small-scale clustering of LAEs near quasars. However, the small field of view of our MUSE observations prevent us from observing large scales where the correlation of LAEs in different parent haloes dominates. This caveat prevents us from interpreting further the correlation length in the framework of bias models.

By stacking all the fields in our sample, our analysis has further shown that the spatially projected alignment of LAEs is remarkably uniform around the central quasar, suggesting that these objects are orbiting in/around relatively massive haloes. Some fields, however, do show clustered LAEs at large projected separations (see e.g. Figure \ref{fig:J033_emitters} for an example), indicating that at least some of our LAE sample could have been accreted through filaments as also found by \citet{Moller01} near a $z\approx3$ quasar.

We also found that the number of LAEs detected in each quasar field decreases with redshift and this quantity is not correlated with the LAE luminosity nor with the quasar luminosity. We have shown that this result can be explained by an increase in the LAEs mass at fixed average \lya\ luminosity as the Universe evolves. This is expected by models of galaxy formation where more massive LAE haloes are accreted onto the quasar host haloes at later times \citep{White91,DeLucia07,DeLucia12}. 

In summary, while we cannot robustly quantify the halo mass of our quasar sample with the data in hand,  the observed kinematics, number density, and spatial distribution  suggest that the population of LAEs we identify resides the near environment of halos with masses in the range $M_{\rm h,QSO} = 10^{12.0-12.5}~{\rm M}_\odot$, a value which is typical for high-redshift quasars.

\begin{figure}
    \centering
    \includegraphics[width=0.5\textwidth]{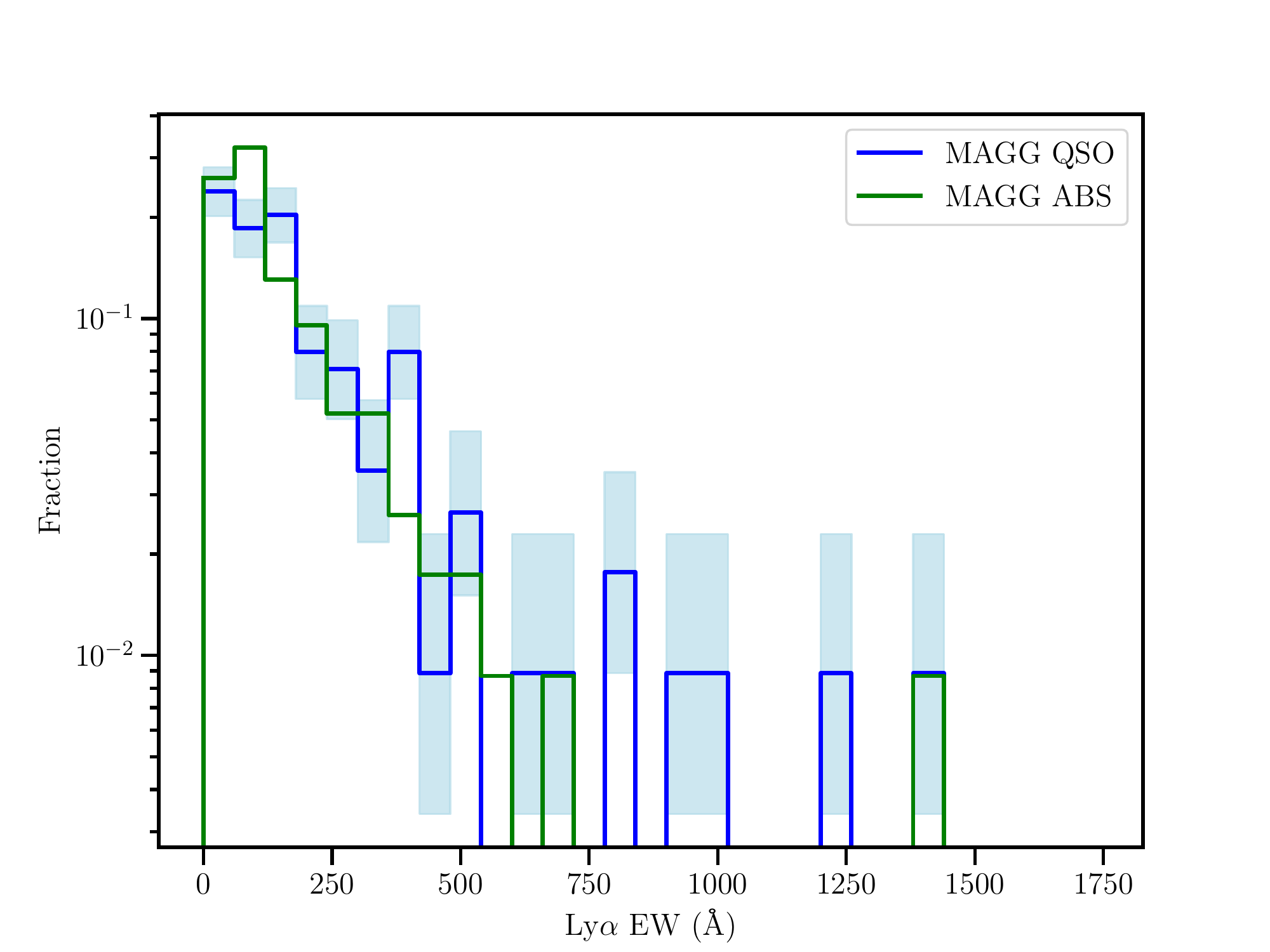}
    \caption{Normalized \lya\ EW distributions in the QSO (blue solid lines) and ABS (green solid lines) samples from our MAGG survey computed following the method presented in \citet{Marino18}. The light blue shaded areas are Binomial uncertainties on the normalized fractions computed from the number counts of the QSO sample. }
    \label{fig:ewlae}
\end{figure}

\begin{figure}
    \centering
    \includegraphics[width=0.5\textwidth]{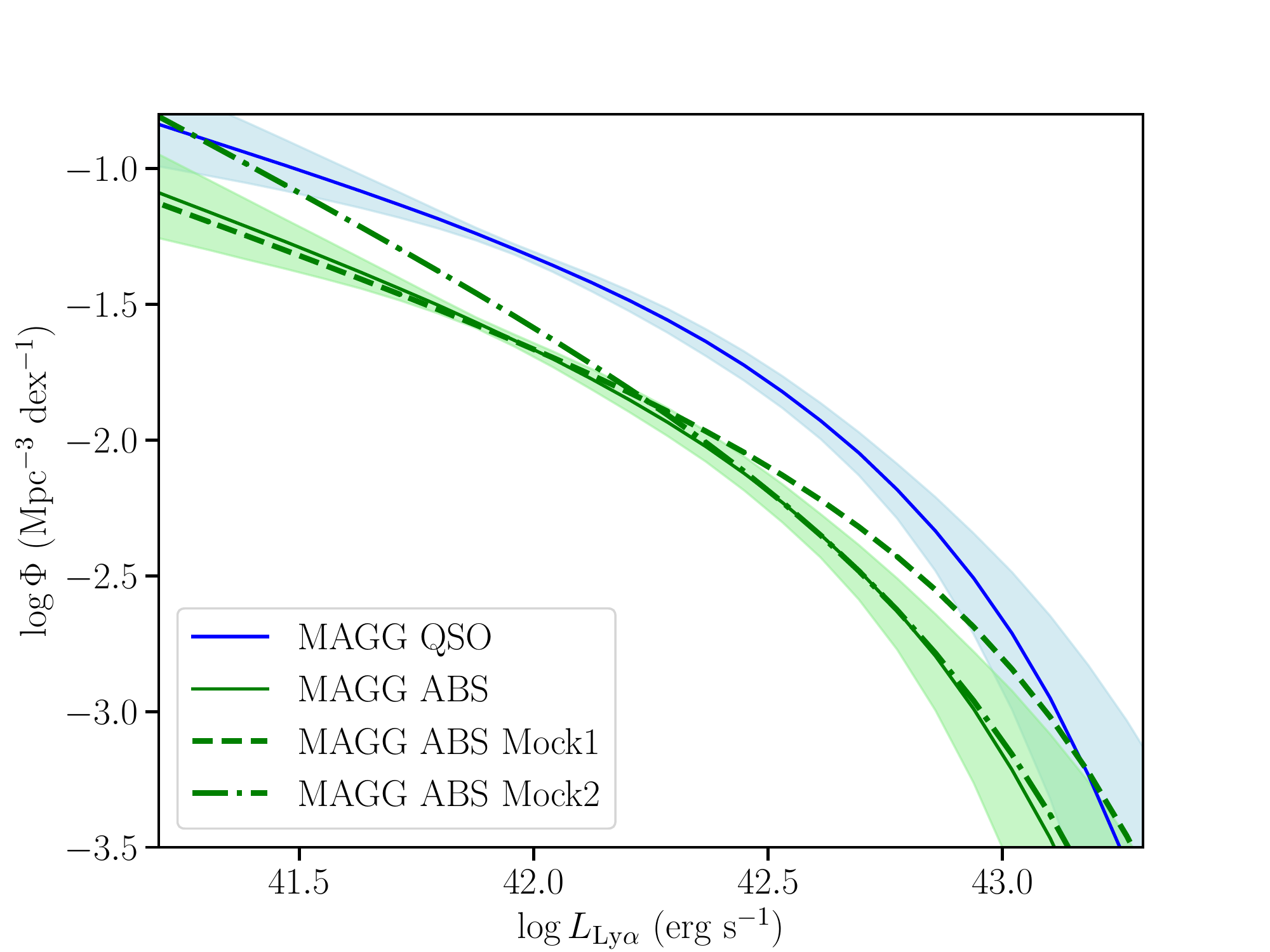}
    \caption{Differential luminosity function of LAEs from the MAGG sample in the quasar environment (QSO, blue solid), and near strong hydrogen absorbers (ABS, green solid). The effect on the ABS luminosity function in presence of a \lya\ luminosity boost due to the quasar ionization field is shown as a dashed line for an excess that depends on the LAE luminosity (Mock 1) and as a dashed-dotted line for a fixed luminosity excess (Mock 2).}
    \label{fig:lumsim}
\end{figure}

\subsection{Effects of the quasar radiation field on LAEs} \label{sec:discussion_radfield}

The analysis of the LAE luminosity function showed a higher density of star forming galaxies in the quasar environment compared to the field and also to the environment of high column-density hydrogen absorbers at the same redshift. \citet{Perez-Rafols18} and \citet{Mackenzie19} have shown that these absorbers are found in more massive haloes compared to the field, possibly extending up to the lower-end of the halo mass range of quasars. 
A second concurrent effect in determining a large number of LAEs near quasars is, however, the boost of \lya\ luminosity due to the quasar ionization field. Indeed, \citet{Cantalupo12} using narrow-band imaging techniques, reported the discovery of nearly 100 \lya\ emitters at $z\approx 2.4$ near a very bright quasar with \Lya\ equivalent width (EW) values so high to make them candidates of star-free dark galaxies. More recently, \citet{Marino18} detected a population of galaxies around six $z>3$ quasars with a higher \lya\ EW  compared to the field emitters, which were also interpreted as low-mass and gas-rich galaxies (also including putative dark galaxies), the emission of which is caused by fluorescence induced by quasar radiation. 

Ultimately, both the parent halo mass (which regulates the intrinsic number of galaxies) and the radiation from the quasars (which regulates the visibility of LAEs either as a boost or a suppression) are responsible in setting the observed number of LAEs. 
An interesting exercise is therefore to assess what is the relative contribution of these two effects. It is however a particularly difficult task to separate the two contributions in absence of tracers that are not affected by radiation (for instance the UV continuum).
Moreover, for the reasons discussed above, it is rather challenging to develop models based from first principle arguments that disentangle these two effects. 

In the spirit of the modelling proposed above for the gas metallicity, we can nevertheless offer some insight into the relevance of quasar boosting by comparing the \lya\ equivalent width (EW) distributions of LAEs in the environment of quasar and high column density absorbers. The EW quantitatively describes the strength of emission features compared to the continuum emission. In case of pure photoionization due to star formation, the EW depends on the gas metallicity and specific star formation rates, but if other processes are boosting the \lya\ line flux this will result in larger EWs (higher fluxes at constant continuum). 
Due to the large fraction of continuum undetected LAEs in  our sample, we follow the method presented in \citet{Marino18} and we compute \lya\ EWs as follows: for continuum detected objects we take the MUSE white-light image flux density in the \lya\ curve of growth aperture, while for continuum undetected objects we take the flux density in an aperture with radius equal to the PSF half width at half maximum (HWHM) at the position of the object, plus the characteristic 1$\sigma$ noise value as a conservative upper limit. In both cases, the continuum flux densities are extrapolated to the wavelength of the \lya\ line, assuming that the monochromatic fluxes are flat in frequency space. Lastly, rest-frame EWs are computed as the ratio between line flux and continuum flux density, and the upper limits for non detected continuum sources become lower limits on the EW. We stress that the EW estimation method is not unique, but it is applied consistently for all LAE samples in our survey. 

In Figure~\ref{fig:ewlae} we show the normalized distributions of LAE EWs in the QSO and ABS samples, noting that lower limits are also included in the distribution. We note in the QSO sample the presence of a handful of emitters above $\approx 750~$\AA, the limit above which nearly no emitters appear in the ABS sample. 
We use a one sided Kolmogorov-Smirnov (KS) test to verify if the empirical cumulative distribution function of the QSO EW distribution is greater than that of the ABS sample, assuming in this case that lower limits are actual values. We find a probability of the null hypothesis $P({\rm KS,null}) = 0.14$, which indicates that, although the QSO sample has more objects at larger EW, this is not highly significant given our sample size. Since the KS test is more sensitive to the centre of the distributions we also test the two samples with an Anderson-darling (AD) test, which instead is more sensitive to differences in the tails. The AD probability of the null hypothesis is $P({\rm AD,null}) = 0.07$, further indicating that we cannot decisively rule out the hypothesis that the two samples are drawn from the same parent distribution.
We note that a significant caveat in this method is given by the large fraction of continuum undetected objects in our LAE samples which correspond to EW limits. Moreover, the mild excess of high EW emitters in the quasar environment is somewhat dependent on the method assumed for the calculation of EWs. For instance, this mild excess disappears if we replace the continuum determination with values computed in apertures matched to the Ly$\alpha$ projected size. Therefore, from this analysis, we cannot conclude that Ly$\alpha$ boosting is confidently detected in this sample. 

The distribution presented in Figure~\ref{fig:ewlae} also adds constraints on the amount of quasar boosting in these fields, ruling out extreme contributions. We must also note that gas photo-evaporation can reduce the  number of high-EW sources around quasars \citep{Kashikawa07, Uchiyama19} eventually leading to suppression of star formation in low mass LAEs. This effect acts in the opposite direction of quasar boosting. The lack of a significant number of continuum detected LAEs in our sample hampers a detailed analysis of the relative contribution of these effects. Sensitive observations in the near-IR with facilities like the Hubble or the James Webb Space Telescopes are required to obtain a direct detection of the LAE continuum for large enough samples to more accurately quantify the various effects of the quasar field on the \lya\ EWs.

In addition to the analysis of the EWs presented above, the LAE luminosity function can offer additional constraints on the effects of quasar ionization boosting. To explicitly show how, we construct simple toy models for the quasar boosting on a population of LAEs. 
The simplicity of our toy models implies they can only be considered as idealized scenarios, highly unlikely to happen in real objects. Moreover, we can only simulate what would happen to a population of objects under these idealized conditions and our analysis does not capture the variable physical conditions of individual LAEs (e.g. gas content, physical distance from the quasar) which could respond differently to the quasar ionization field. Rather, these models are helpful in illustrating a key point: whether radiation boosting is the dominant factor in shaping the luminosity function, that is associated in a transformation rather than a simple re-normalization of the luminosity function.    

In this analysis, we model the boost in two simple but different ways. We assume either that the quasar ionization contributes to a \lya\ luminosity excess that is equal to 20\% of the luminosity of each LAE (Mock 1), or that it adds a fixed quantity to the \lya\ luminosity of all emitters equal to 5\% of $L^*$ (Mock 2). For each of these scenarios, we draw a random sample from the ABS luminosity function, we apply the boost to the \lya\ luminosity and we fit a Schechter function by including the survey selection function as described above. 
We note that the numerical values reported for the boost factors are purely indicative of the models we present here; we have experimented with a range of values finding the same qualitative trends. 

The effect of the two toy models on the ABS luminosity function is shown in Figure~\ref{fig:lumsim}. The dashed green line shows the effect of a luminosity excess that is proportional to the \lya\ luminosity of LAEs (Mock 1): in this case, the luminosity function is affected in the value of $L^*$ which moves to higher luminosity, but not in the faint-end slope nor in the absolute normalization. Conversely, a fixed \lya\ luminosity excess (Mock 2) increases the faint-end slope without an appreciable effect on the normalization or $L^*$. \citet{Cantalupo12} also studied the effects of quasar boosting on the LF of LAE finding that a steepening of the faint end slope (our Mock 2) would be the clearest signature of fluorescent emission on the LF. In both our mocks, we are unable to reproduce the luminosity function near quasars by making simple assumptions on the quasar ionization effect. Indeed, the ABS and QSO LF have a very consistent shape with the only exception of the normalization. 

In line with all the arguments presented above, this additional piece of evidence is in support of the idea that the distribution of LAEs around quasars must be intrinsically different, and that quasar boosting is not the only effect at play. Indeed, further support for a non overwhelmingly dominant effect of the quasar radiation field comes from our analysis of the properties of the CGM of the LAE sample and from the lack of correlation between the LAE luminosity and the absolute magnitude of the quasar in their field. Moreover, the stacked LAE spectrum for the quasar environment does not show prominent emission lines originating from a hard AGN ionization field (e.g. \CIV) and it is consistent with that of the ABS environment (which does not have a central ionizing source). Moreover, we do not detect any clear asymmetry (down to $3.2\times 10^{-19} (2\sigma)$ \sbline) when we orient and stack the LAE sample in the direction of the quasars (see Figure~\ref{fig:lae_nbstack}).

Collectively, all these pieces of evidence point to a scenario where quasars are hosted in massive haloes (possibly more massive than strong hydrogen absorbers) with a population of LAEs that is intriniscally different and not just affected by the quasar ionization field. We must note however that in the literature it is possible to find examples of structures that are able to ionize large gas bubbles around them. \citet{Umehata19} found \lya\ radiation extending on the Mpc scale in the SSA22 field at $z\sim 3.1$ where a significant number of highly star-forming galaxies and AGN are found to be powering the emission. More recently, \citet{Mukae20} used the quasar tomographic technique coupled with the detection of LAEs to uncover an extreme overdensity of six quasars and four LAEs at $z\sim 2.1$. This overdensity correlates with a 40 Mpc wide \HI\ underdensity indicating that the ionizing radiation of the quasar has created a large and ionized gas bubble. These environments, however, are rare and likely more extreme than our MAGG fields and the ionization conditions found there might not necessarily apply to the average quasar population.

\section{Conclusions} \label{sec:conclusions}

In this work, we have presented a complete analysis of the environment of $z=3-4.5$ bright quasars using data from the MUSE Analysis of Gas around Galaxies (MAGG) survey. MAGG is a VLT large programme covering 28 quasar fields with medium-deep integral field observations with the MUSE instrument. We have characterized the effects of 27 quasars (one field is gravitationally lensed and therefore excluded from this work) on their surroundings by studying the properties of extended ionized gas nebulae and the distribution, luminosity and morphology of LAEs in the quasar host haloes.

Extended ionized gas nebulae are detected around each MAGG quasar, consistently with what was found in the literature for other quasar surveys. We derived individual and stacked radial profiles of the \lya\ emission from the nebulae finding little or no evolution compared to previous MUSE surveys over the redshift range $z\approx 3-6$. A significant evolution towards higher surface brightness is instead found when we compare our MAGG data at $z\sim 3.8$ with the \citet{Cai19} survey at $z\sim2$. At lower redshift, it is possible that the growth of haloes shock heats the CGM gas reducing the cold gas mass and leading to a fainter \lya\ profile.

Thanks to the depth of the MAGG data we have also searched for extended emission from metal lines in the nebulae. By stacking the 27 MUSE datacubes we have  detected \CIV\ extended emission and placed a strong upper limit on the \HeII\ emission. The low \HeII/\Lya\ ratio in our larger sample, consistent with the results of \citet{Cantalupo19} for the Slug nebula at $z\sim 2.3$, suggests that the CGM of bright quasars might have a very broad and clumpy density distribution also at the higher redshift probed by our sample. 
Given the likely multi-phase and clumpy nature of the CGM that is not properly captured in current models, it is very difficult to obtain a metallicity estimate from the \CIV\ emission. The detection of \CIV\ indicates that the extended gas is somewhat metal enriched but with the current data it is difficult to formulate more firm conclusions.



One distinctive feature of the MAGG program is its ability to map the population of LAEs in the vicinity of the quasars, thanks to the depth of the observations. We have performed a detailed analysis of the accuracy of the detection algorithms and of the completeness of the catalogue extraction using mock sources, and we found a sample of 113 LAE in our quasar fields. Their redshift and spatial clustering suggest that these LAEs are star-forming galaxies orbiting the quasar halo potential.

We then built the luminosity function of LAEs in the quasar environment and compared it to that in the environment of strong hydrogen absorbers and to the field. We found more LAEs around quasars although the shape of the luminosity function remains consistent across all the three environments we studied. We used simple toy models to show that a boost in the \lya\ luminosity due to the quasar ionization field is unable to explain in itself the observed offset in luminosity function between the quasar and the absorbers environment. 
We also observed a lack of correlation between the quasar absolute magnitude and the average (or highest) luminosity of LAEs in the same field. Lastly, we did not detect a significant asymmetry in the morphology of LAEs if they are stacked after orienting them such that the quasar always points in the same direction.

All these results suggest that the quasars are
hosted by haloes in the mass range $ \approx 10^{12.0}-10^{12.5}~\rm M_\odot$, and are surrounded by a larger population of LAEs compared to the environment of strong absorbers. This excess is likely caused by the denser local environment, and not exclusively by the Ly$\alpha$ boosting induced by the quasar radiation. Moreover, our analysis places sufficient constraints to rule out extreme contributions of the quasar radiation field to the observed LAE properties. Future observations covering a larger footprint around the quasars as well as non-fluorescent tracers such as the UV continuum will be critical to reveal the full extent of LAE clustering around high-redshift quasars, and to disentangle the concurrent effects of clustering and quasar boosting in shaping the observed distribution of LAEs.

\section*{Acknowledgements}
We thank the anonymous referee whose comments have improved the quality of the manuscript.
This project has received funding from the European Research Council (ERC) under the European Union's Horizon 2020 research and innovation programme (grant agreement No 757535) and by Fondazione Cariplo (grant No 2018-2329). 
SC gratefully acknowledges support from the Swiss National Science Foundation grant PP00P2\_190092 and from the European Research Council (ERC) under the European Union’s Horizon 2020 research and innovation programme grant agreement No 864361. MTM thanks the Australian Research Council for \textsl{Discovery Project} grants DP130100568, DP170103470 and DP190100417 which supported this work. This work is based on observations collected at the European Organisation for Astronomical Research in the Southern Hemisphere under ESO programme IDs 
197.A-0384, 
065.O-0299,
067.A-0022,
068.A-0461,
068.A-0492,
068.A-0600,
068.B-0115,
069.A-0613,
071.A-0067,
071.A-0114,
073.A-0071,
073.A-0653,
073.B-0787,
074.A-0306,
075.A-0464,
077.A-0166,
080.A-0482,
083.A-0042,
091.A-0833,
092.A-0011,
093.A-0575,
094.A-0280,
094.A-0131,
094.A-0585,
095.A-0200,
096.A-0937,
097.A-0089,
099.A-0159,
166.A-0106,
189.A-0424.
This work used the DiRAC Data Centric system at Durham University, operated by the Institute for Computational Cosmology on behalf of the STFC DiRAC HPC Facility (www.dirac.ac.uk). This equipment was funded by BIS National E-infrastructure capital grant ST/K00042X/1, STFC capital grants ST/H008519/1 and ST/K00087X/1, STFC DiRAC Operations grant ST/K003267/1 and Durham University. DiRAC is part of the National E-Infrastructure. This research made use of Astropy \citep{Astropy-Collaboration13}. 

\section*{Data Availability}
The data used in this work are available from the ESO Science Archive Facility \url{https://archive.eso.org/}, while the codes used in this work are available at \url{http://www.michelefumagalli.com/codes.html} The {\sc CubExtractor} package is available from Sebastiano Cantalupo upon request.
  



\bibliographystyle{mnras}







\bsp 
\label{lastpage}
\end{document}